\documentclass[]{JFM-FLM_Au}

\usepackage{float}
\usepackage[section]{placeins}
\usepackage{xcolor}

\lefttitle{Reverte-Badillo et al.}
\righttitle{Journal of Fluid Mechanics}

\title{What explainable deep learning reveals about wall-bounded turbulence: insights from conditional surprisal}

\author{Belén Reverte-Badillo\aff{1}, Andrés Cremades\aff{1}, Ricardo Vinuesa\aff{2} \and Sergio Hoyas\aff{1}}

\affiliation{\aff{1}Instituto Universitario de Matemática Pura y Aplicada, Universitat Politècnica de València, 46022 Valencia, Spain
\aff{2}Department of Aerospace Engineering, University of Michigan, Ann Arbor, MI 48109, USA}

\corresau{Sergio Hoyas, email: sergio.hoyas@upv.es}

\begin{document}
\maketitle

\begin{abstract}
Explainable deep learning (XDL) has identified regions of wall turbulence important for prediction that are only partly captured by classical coherent-structure criteria. We examine turbulent channel flow at friction Reynolds numbers $Re_\tau\simeq125$ and $550$ by comparing SHapley Additive exPlanations (SHAP) with conditional surprisal, which quantifies how unexpected an observed velocity transition is given its initial state. Although SHAP selection favours energetic fluid and largely overlaps classical masks, selected fluid outside these masks is less energetic but exhibits higher surprisal. Near the wall, at comparable fluctuation energy and wall distance, high-surprisal voxels outside classical masks are approximately 11 and 5 times more likely to be selected by SHAP than low-surprisal voxels, respectively. Within classical masks, the factor is about 1.3. SHAP-selected structures containing paired high- and low-speed regions exhibit surprisal maxima near their internal $u=0$ interfaces, where transverse velocity fluctuations remain nonzero. These interfaces contain vortical and SHAP-only regions, with the latter accounting for more than half of their reconstructed area. In part of the buffer layer, the SHAP-only portions have higher mean surprisal than the vortical portions. However, comparison with generic $u=0$ surfaces indicates that much of the raw interfacial surprisal excess is associated with surrounding fluctuation energy; after energy matching, the equally weighted surface comparison is statistically compatible with unity on the retained common support. Together, these results show how state-conditioned velocity transitions complement instantaneous structure-identification criteria in characterizing SHAP-selected flow.
\end{abstract}


\section{Introduction}

What makes a region of a turbulent flow important for its subsequent evolution? In wall-bounded turbulence, coherent structures provide a physical description of motions that transport momentum and sustain turbulent activity~\citep{Jimenez2018}. Regions identified by their instantaneous intensity, however, do not necessarily coincide with those emphasized by a model predicting the flow. Recent studies using explainable deep learning have exposed this difference, identifying regions that overlap only partly with classical coherent structures \citep{Cremades2024,Cremades2025,hoyas2025deep}. These findings leave a physical question open: what distinguishes the selected regions, particularly the fluid that conventional criteria do not capture?

Classical descriptions of wall turbulence have developed from observations of near-wall streaks and bursting \citep{Kline1967,KimKlineReynolds1971} into a broader account of coherent motions and their organization across scales \citep{Robinson1991,Adrian2007,Jimenez2018}. Intense Reynolds-stress events \citep{WallaceEckelmannBrodkey1972,LuWillmarth1973,LozanoDuranFloresJimenez2012}, streaks \citep{SmithMetzler1983} and vortical structures identified from the velocity-gradient tensor \citep{Chong1990,JeongHussain1995} describe different aspects of momentum transfer and the regeneration of turbulence. Their interactions underpin the near-wall self-sustaining cycle, in which vortical motions generate streaks and the subsequent evolution of the streaks helps regenerate the vortices \citep{HamiltonKimWaleffe1995,JimenezPinelli1999,Waleffe1997,SchoppaHussain2002}. Tracking these structures has also established how they form, merge, split, and decay \citep{LozanoDuranJimenez2014}. These descriptions associate recognizable flow configurations with turbulent activity. However, predictive attribution and state-conditioned velocity transitions offer a complementary perspective on these flow configurations.

Neural networks trained on simulation data offer another way to examine the evolution of turbulence \citep{Srinivasan2019,Brunton2020,Guastoni2021}. Their predictions can be analysed using SHapley Additive exPlanations (SHAP) \citep{Shapley1953,LundbergLee2017}, which assigns contributions to input features using the Shapley value of cooperative game theory. \citet{Cremades2024} applied SHAP to intense Reynolds-stress structures in turbulent channel flow, explaining the mean-squared error of a network forecasting the velocity field. They found that the importance assigned to a structure was not simply determined by its contribution to Reynolds shear stress. \citet{Cremades2025} extended the analysis to individual grid points and defined connected SHAP structures by thresholding the attribution magnitude. These structures overlapped only partly with Q events, streaks, and vortical structures. The approach has also been applied to two-dimensional flow sections \citep{MolinaCasino2025} and reviewed for fluid dynamics and heat transfer \citep{CremadesReview2025}. Its practical relevance is supported by the use of attribution-based rewards for turbulent drag reduction \citep{Beneitez2025}. Together, these studies identify flow regions worth examining, but leave their physical interpretation incomplete.

The spatial organization of these regions provides a further clue. In an adverse-pressure-gradient boundary layer over a wing, \citet{MolinaCasino2026} identified SHAP structures containing adjacent high- and low-speed regions. In their ensemble composites, attribution magnitude was concentrated near the internal $u=0$ surface separating the two, where $u$ denotes the streamwise velocity fluctuation. Interfaces between high- and low-speed fluid are familiar features of wall turbulence. Sweeps and ejections form spatially organized pairs near streak boundaries in channels \citep{LozanoDuranFloresJimenez2012}, while zones of nearly uniform streamwise momentum in boundary layers are separated by thin shear layers \citep{MeinhartAdrian1995,deSilva2016}. Near-wall shear layers have also been associated with turbulence production \citep{Johansson1991} and high-amplitude wall-pressure events \citep{GhaemiScarano2013}. This suggests a connection between SHAP-selected pairs and established flow organization. 

The distinction between instantaneous intensity and subsequent evolution also appears in studies based on controlled perturbations. \citet{Jimenez2018machine} measured the significance of a flow subvolume through its influence on the later flow in two-dimensional decaying turbulence. In three-dimensional isotropic turbulence, kinetic energy identifies regions of large absolute perturbation growth much more effectively than regions of large relative amplification \citep{EncinarJimenez2023}. In channels, \citet{OsawaJimenez2024} associated causally significant regions with sweeps and, near the wall, with high-speed fluid overtaking low-speed fluid. Information-theoretic methods address related questions from observations, including interactions between energy-containing eddies \citep{LozanoDuranBaeEncinar2020} and the information that velocity components carry about future wall shear stress \citep{ArranzLozanoDuran2024}. SHAP values quantify the contribution of each input to a trained model's prediction error; statistical associations with these attributions do not establish the causal influence that would be measured by perturbing the flow. Nevertheless, analyses of simplified dynamical systems and applications to flow control have demonstrated an association between SHAP attributions and causal surrogates (see the Supplementary Information of \citet{Cremades2025}).

Here, we seek a physical interpretation of SHAP-selected regions by examining the velocity transitions that occur within them. For a given initial velocity, some observed changes are predictable, whereas others are unusual. Conditional surprisal quantifies this distinction through $S=-\log\widetilde f(Y\mid X)$, where $X$ is the initial streamwise velocity, $Y$ is its value at a convectively displaced location in the subsequent velocity field, five viscous time units later, and $\widetilde f$ is an estimated conditional probability density. This is the information content of an observed outcome conditioned on its initial state \citep{Shannon1948,CoverThomas2005}. Following the approximate downstream convection of the flow reduces the contribution of structures simply passing a fixed observation point. The resulting quantity describes how unexpected the observed velocity transition is in that frame. It is calculated directly from the velocity fields and can therefore be compared with SHAP without using the forecasting network to define it.

Conditional surprisal is closely related to the magnitude of the convected velocity increment, but also accounts for the initial state from which the change occurs. It thus provides a statistical description of local velocity evolution that can be compared between regions of similar instantaneous fluctuation energy. Unlike conditional entropy, which averages over possible outcomes, surprisal assigns a value to each observed transition \citep{CoverThomas2005,LozanoDuranArranz2022}. It does not measure a network's prediction error or the intrinsic unpredictability of the complete flow. SHAP attribution and surprisal nevertheless both involve the subsequent velocity field, making their relationship relevant to the physical interpretation of the regions identified in the earlier SHAP studies.

We examine this relationship in direct numerical simulations of turbulent channel flow at $Re_\tau\simeq125$ and $550$, using the forecasting networks and attribution fields of \citet{Cremades2025}. The analysis addresses two aspects of the selected flow. First, we compare fluid inside and outside classical coherent-structure masks to determine how fluctuation energy, velocity composition and conditional surprisal distinguish SHAP selection. Comparisons at matched energy and wall distance separate differences in instantaneous intensity from differences in the observed transitions. Second, we identify SHAP structures containing paired high- and low-speed regions and examine the distribution of surprisal around their internal $u=0$ interfaces, together with the velocity content and classical structures found on those surfaces. The two analyses concern selected portions of the flow and complete connected structures, respectively. Fluid outside a classical mask need not form a separate structure. The recurrence of the main findings at both Reynolds numbers supports the interpretation that SHAP values identify physical phenomena rather than merely model-specific patterns.

Concerning the article structure, section~\ref{sec:methodology} describes the databases, structure definitions and statistical methods. The results are presented in \S\ref{sec:results}, followed by their physical interpretation in \S\ref{sec:discussion} and the conclusions in \S\ref{sec:conclusions}. The appendices document estimator validation and the sensitivity of the results to the observation frame, statistical conditioning and interface reconstruction.

\section{Methodology}
\label{sec:methodology}

\subsection{Flow configuration and numerical database}
\label{sec:dns}

We analyse two direct numerical simulation (DNS) databases of turbulent plane-channel flow from \citet{Cremades2025}, at friction Reynolds numbers $Re_\tau=u_\tau h/\nu\simeq125$ and $550$, where $u_\tau$ is the friction velocity, $h$ the channel half-height and $\nu$ the kinematic viscosity. The simulations were performed with LISO \citep{hoyas2006scaling,Hoyas2022}, a pseudo-spectral solver based on \citet{KimMoinMoser1987} that employs a seven-point compact finite-difference scheme in the wall-normal direction \citep{Lele1992}. Computational domains and grid resolutions are summarized in \autoref{tab:dns_parameters}. Throughout, superscript $+$ indicates normalization in wall units, with $y^+=yu_\tau/\nu$ denoting distance from the nearest wall and $t^+=tu_\tau^2/\nu$.

\begin{table}
    \centering
    \begin{tabular}{ccccccc}
        $Re_\tau$ & $(L_x,L_y,L_z)/h$ & $N_x\times N_y\times N_z$
        & $\Delta x^+$ & $\Delta z^+$ & $\Delta y^+_{\min}$ & $\Delta y^+_{\max}$ \\
        \hline
        $125$ & $(8\pi,2,3\pi)$ & $384\times201\times288$
        & $8.18$ & $4.09$ & $0.34$ & $1.61$ \\
        $550$ & $(2\pi,2,\pi)$ & $384\times251\times384$
        & $8.8$ & $4.4$ & $0.78$ & $5.77$ \\
    \end{tabular}
    \caption{Computational domains and grid resolutions of the DNS databases. Grid spacings are approximate and expressed in wall units. Both simulations are periodic in the streamwise and spanwise directions and employ non-uniform wall-normal grids.}
    \label{tab:dns_parameters}
\end{table}

Conditional surprisal is computed from approximately $8000$ consecutive pairs of three-dimensional velocity fields at each Reynolds number, separated by $\Delta t^+=5$. Comparisons with SHAP are restricted to snapshots with consistent attribution maps, reference profiles, and masks \citep{Cremades2025}, yielding approximately $8000$ pairs at $Re_\tau\simeq125$ and $3400$ at $Re_\tau\simeq550$. All surprisal calculations use the original DNS fields rather than neural-network predictions. Smaller subsets employed in sensitivity tests are specified where appropriate.

The instantaneous velocity is denoted by $\boldsymbol U=(U,V,W)$, with mean $\overline{\boldsymbol U}(y)$ and fluctuations $\boldsymbol u=(u,v,w)=\boldsymbol U-\overline{\boldsymbol U}$. Their root-mean-square values are $u'(y)=\overline{u^2}^{1/2}$, $v'(y)=\overline{v^2}^{1/2}$ and $w'(y)=\overline{w^2}^{1/2}$. We define the local energy measure $E(\boldsymbol{x},t)=u^2+v^2+w^2$, proportional to the turbulent kinetic energy per unit mass, for subsequent relative-energy comparisons.

\subsection{SHAP-selected and classical coherent structures}
\label{sec:structures}

We use the precomputed SHAP attribution fields and coherent-structure masks of \citet{Cremades2025}. These attributions were obtained from three-dimensional convolutional neural networks with U-net architectures \citep{Ronneberger2015}, trained to predict velocity fluctuations five viscous time units ahead. At each Reynolds number, $10\,000$ velocity fields were divided into $80\,\%$ for training and $20\,\%$ for validation. The reported relative prediction errors for the streamwise, wall-normal and spanwise velocity components were $1.15\,\%$, $0.98\,\%$ and $1.08\,\%$ at $Re_\tau\simeq125$, and $0.74\,\%$, $1.15\,\%$ and $0.88\,\%$ at $Re_\tau\simeq550$. The attribution fields analysed here correspond to snapshots outside the networks' training periods.

Each component of the velocity fluctuations was normalized using fixed global minimum and maximum values, determined separately for each component over the entire spatial domain and the snapshots used to establish the normalization:
\begin{equation}
    \widetilde q =
    \frac{q-q_{\min}}{q_{\max}-q_{\min}},
    \qquad q\in\{u,v,w\}.
    \label{eq:minmax_scaling}
\end{equation}
The extrema were kept fixed across all snapshots for each Reynolds number. The zero-fluctuation reference field was normalized using the same extrema, giving $\widetilde q^{\mathrm{ref}}=
-q_{\min}/(q_{\max}-q_{\min})$, which is generally nonzero.

The scalar quantity explained by SHAP is the scaled mean-squared prediction error of the normalized velocity field, rather than the predicted field itself. It is defined as
\begin{equation}
    \mathcal M(\boldsymbol{x};\boldsymbol{y})=\frac{10^4}{3N}\sum_{i=1}^{N}\sum_{c\in\{u,v,w\}}\left[F_c(\boldsymbol{x})_i-y_{c,i}\right]^2,
    \label{eq:prediction_mse}
\end{equation}
where $F_c(\boldsymbol{x})_i$ is the normalized velocity predicted by the U-net for component $c$ at grid point $i$, $\boldsymbol{x}$ is the normalized input field, $\boldsymbol{y}$ is the corresponding normalized DNS field at the forecast time, and $N=N_x\times N_y\times N_z$ is the total number of grid points. The squared errors are averaged over all grid points and velocity components without volume weighting, and the result is multiplied by $10^4$.

Following \citet{Cremades2025}, attributions were computed using the expected-gradients framework \citep{Erion2021}, with a single fixed zero-fluctuation reference field and 200 randomly sampled interpolation points. With this fixed reference, the method reduces to integrated gradients, which are approximated using Monte Carlo sampling. The attribution of input feature $i$, corresponding to one velocity component at one grid point, is

\begin{equation}
    \phi_i=
    \bigl(x_i-x_i^{\mathrm{ref}}\bigr)
    \mathbb E_{\alpha\sim\mathcal U(0,1)}
    \left[
    \frac{\partial\mathcal M\bigl(\boldsymbol{x}^{\mathrm{ref}}+\alpha(\boldsymbol{x}-\boldsymbol{x}^{\mathrm{ref}});\boldsymbol{y}\bigr)}{\partial x_i}
    \right].
    \label{eq:expected_gradients}
\end{equation}

Here, $\boldsymbol{x}^{\mathrm{ref}}$ is the normalized zero-fluctuation reference. During interpolation between this reference and the input field, the DNS target $\boldsymbol{y}$ remains fixed. Since the same min--max scaling is applied to both the input and reference, $x_i-x_i^{\mathrm{ref}}$ is proportional to the corresponding physical velocity fluctuation. Consequently, the attribution of each component vanishes wherever that fluctuation is zero. These attributions describe the contribution of individual input features to the prediction error, rather than their direct causal influence on the flow.

The three attribution components are combined into the magnitude
\begin{equation}
    \Phi(\boldsymbol{x},t)
    =\left(\phi_u^2+\phi_v^2+\phi_w^2\right)^{1/2}.
    \label{eq:shap_magnitude}
\end{equation}

This magnitude discards the signs of the individual attributions, allowing regions with large contributions to the prediction error to be identified regardless of their sign. An analysis of the signed attributions shows that $\phi_u+\phi_v+\phi_w$ is negative in more than $98\,\%$ of SHAP-only voxels at both Reynolds numbers, with negligible cancellation in the accumulated contributions. Thus, the selected inputs predominantly contribute to reducing the network's prediction error relative to the zero-fluctuation reference. The SHAP-selected mask $\mathcal H$ is then defined by thresholding the attribution magnitude:

\begin{equation}
    \mathcal H(t)=
    \left\{
    \boldsymbol{x}:
    \Phi(\boldsymbol{x},t)>
    H\left[
    \overline{\phi_u^2}(y)+
    \overline{\phi_v^2}(y)+
    \overline{\phi_w^2}(y)
    \right]^{1/2}
    \right\},
    \label{eq:shap_threshold}
\end{equation}
where the overbar denotes averaging over the homogeneous directions and the attribution ensemble. We use the original percolation thresholds, $H=2$ at $Re_\tau\simeq125$ and $H=1.95$ at $Re_\tau\simeq550$, together with the corresponding reference profiles supplied with the attributions. The original masks are retained without imposing a minimum-volume filter on the voxel-wise comparisons. Connected SHAP structures are identified with six-neighbour connectivity and periodicity in the homogeneous directions; the volume filter introduced in \S\ref{sec:pairs} applies only to the analysis of complete structures.

Ensemble composites use the snapshot-normalized magnitude $\Phi_g=\Phi/\langle\Phi\rangle_{xyz}(t)$, which differs from the ensemble reference used to define the SHAP mask.

The classical coherent-structure masks follow \citet{Cremades2025}. Intense Reynolds-stress events, or Q events ($\mathcal Q$), satisfy
\begin{equation}
    |uv|>\beta\,u'(y)v'(y),
    \qquad \beta=1.75,
\end{equation}
and include all four quadrants. Streaks ($\mathcal L$) are defined by
\begin{equation}
    (u^2+w^2)^{1/2}>\alpha_s u_\tau,
    \qquad \alpha_s=4,
\end{equation}
and include both high- and low-speed fluid. Vortical structures ($\mathcal V$) are identified using the velocity-gradient discriminant criterion of \citet{Chong1990}, thresholded at each height at $\gamma=0.02$ times the standard deviation of the discriminant. These thresholds correspond to the percolation criteria of \citet{Cremades2025}. The sensitivity of the results to these thresholds is examined in Appendix~\ref{app:threshold_sensitivity}.

The union of the classical masks is
\begin{equation}
    \mathcal C=\mathcal Q\cup\mathcal L\cup\mathcal V.
    \label{eq:classical_union}
\end{equation}
For direct comparison with SHAP, the flow is divided into four mutually exclusive populations:
\begin{subequations}\label{eq:categories}
\begin{align}
    \mathcal R_{\mathrm{overlap}}
    &=\mathcal H\cap\mathcal C,\\
    \mathcal R_{\mathrm{SHAP\text{-}only}}
    &=\mathcal H\setminus\mathcal C,\\
    \mathcal R_{\mathrm{classical\text{-}only}}
    &=\mathcal C\setminus\mathcal H,\\
    \mathcal R_{\mathrm{background}}
    &=(\mathcal H\cup\mathcal C)^{\mathrm c}.
\end{align}
\end{subequations}
These populations are disjoint and cover the analysis domain. The term \emph{SHAP-only} refers to selected voxels outside every classical mask; it does not imply that those voxels form independent connected SHAP structures.

\subsection{Convectively aligned velocity observations}
\label{sec:convective_frame}

Temporal changes in turbulent velocity reflect both the downstream advection of coherent structures and changes in their internal organization. At a fixed spatial location, the passage of a structure can produce substantial velocity variations even if its internal configuration remains approximately unchanged. We therefore use a convectively displaced observation point to reduce the contribution of advection to the measured velocity change, rather than interpreting every fixed-point temporal variation as local evolution.

The instantaneous streamwise velocity at $(x,y,z,t_s)$ is paired with its value at a downstream location in the subsequent stored field:
\begin{equation}
    X=U(x,y,z,t_s),
    \qquad
    Y=U(x+\ell(y),y,z,t_{s+1}),
    \label{eq:surprisal_variables}
\end{equation}
where $\ell(y)=U_{\mathrm{conv}}(y)\Delta t$ and $\Delta t$ is the interval between the stored fields.

The convection-velocity profile is motivated by observations that turbulent fluctuations propagate approximately with the local mean velocity away from the wall, while their propagation velocity approaches a finite, nearly constant value in the near-wall region \citep{KimHussain1993}. The convection velocity also depends on the spatial scales of turbulent motions, with near-wall streaks exhibiting characteristic propagation velocities of approximately $11u_\tau$ \citep{delAlamoJimenez2009}.

We therefore prescribe
\begin{equation}
    U_{\mathrm{conv}}^+(y)
    =
    \max\left[
    \overline U^+(y),
    \overline U^+(y^+=12)
    \right],
    \qquad
    \ell^+(y)=U_{\mathrm{conv}}^+(y)\Delta t^+.
    \label{eq:convective_velocity}
\end{equation}
This uses the local mean velocity away from the wall and maintains a finite convection speed below $y^+=12$. The reference height provides a practical approximation to the near-wall propagation plateau, rather than a universal convection velocity for all turbulent scales. Its consistency with convection velocities inferred from the DNS and the sensitivity of the results to alternative observers, with the prescribed observer denoted $S_3$ in that comparison, are examined in Appendix~\ref{app:robustness}.

Because $\ell(y)$ is generally not an integer multiple of the streamwise grid spacing, the displaced velocity is evaluated using the Fourier shift theorem. Each streamwise Fourier coefficient is multiplied by $\exp[ik_x\ell(y)]$, allowing fractional translation within the periodic domain without rounding the displacement to the nearest grid point. No displacement is applied in the wall-normal or spanwise directions.

This construction follows a prescribed approximation to the mean downstream propagation of turbulent motions, not a material or Lagrangian trajectory. It does not remove scale-dependent or fluctuating advection. Consequently, the conditional surprisal defined below measures unexpected velocity changes relative to this specified observer, rather than an observer-independent property of flow unpredictability.

\subsection{Regularized estimation of local conditional surprisal}
\label{sec:surprisal}

To estimate the conditional density required to compute
surprisal, we use the observed pairs of initial and
convectively displaced streamwise velocities $(X,Y)$.
For each wall-parallel DNS plane at fixed $y$ and each
snapshot pair, we use one velocity pair per mesh point
in the two homogeneous directions, giving
$N_x \times N_z=384\times288=110\,592$ observations at
$Re_\tau\simeq125$ and $384\times384=147\,456$ at
$Re_\tau\simeq550$. The joint distribution is estimated
independently for each plane and snapshot pair,
without temporal pooling. The wall planes, where the
no-slip condition makes the streamwise velocity
identically zero, are excluded.

\begin{figure}
    \centering
    \includegraphics[width=\textwidth]{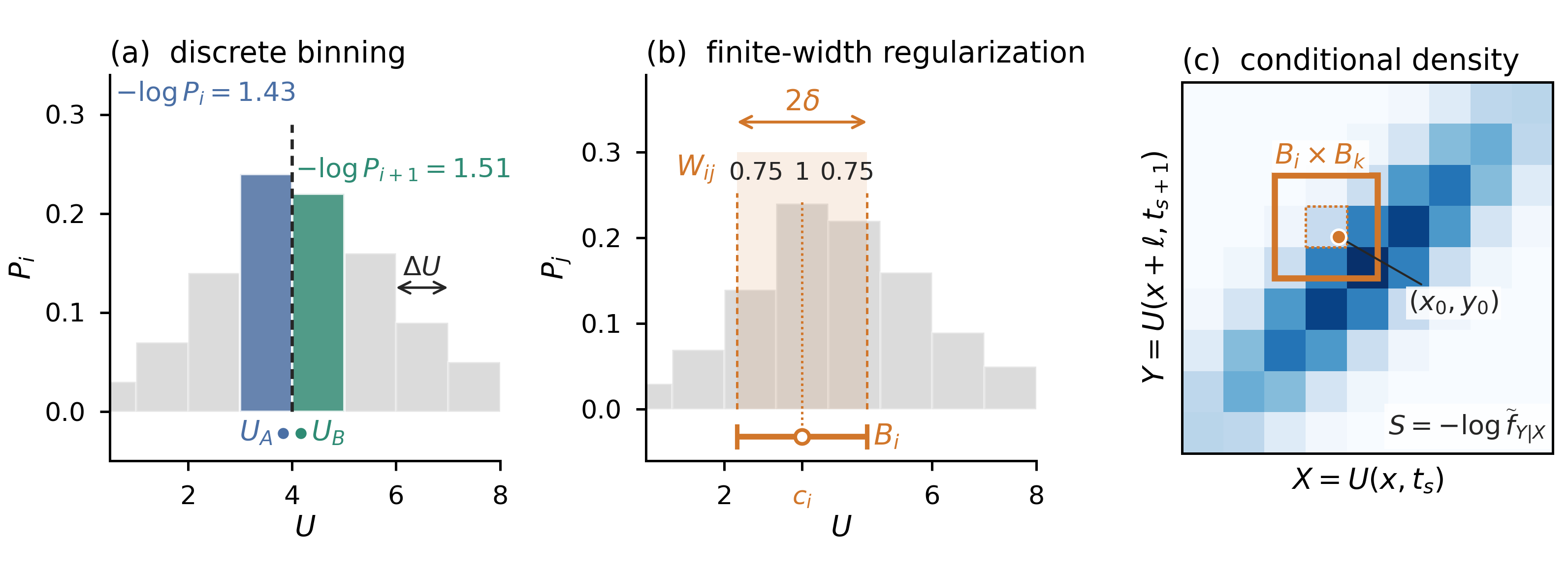}
    \caption{Schematic of the conditional-surprisal estimator. (a) Discretization of continuous velocity observations into histogram bins, illustrating sensitivity to bin boundaries. (b) Finite-width regularization using a window $B_i=[c_i-\delta,c_i+\delta]$, with fractional contributions from neighbouring bins determined by their overlap weights $W_{ij}$. (c) Regularized joint distribution of the initial velocity $X=U(\boldsymbol{x},t_s)$ and the convectively displaced velocity $Y=U(\boldsymbol{x}+\boldsymbol{\ell},t_{s+1})$, showing the window used to estimate the conditional density and surprisal of an observed pair. Values are illustrative and do not represent DNS measurements.}
    \label{fig:surprisal_estimator}
\end{figure}

Because velocity is a continuous variable, the probability of an exact value is not appropriate for measuring how likely an observed transition is. A direct histogram-density estimator can be sensitive to the occupancy of individual joint bins and to the position of bin boundaries. To reduce this sensitivity, we use finite-width regularization windows in velocity space, illustrated in figure~\ref{fig:surprisal_estimator}.

For each plane and snapshot pair, a common velocity interval $[\min(X\cup Y),\max(X\cup Y)]$ is constructed using the initial and convectively displaced velocities. The observations are expressed in the outer velocity units of the simulations, $U_r=Re\,\nu/h$, with $Re=2100$ at $Re_\tau\simeq125$ and $Re=11180$ at $Re_\tau\simeq550$. The interval is divided into $N_b=120$ equal-width bins $I_i$, with centres $c_i$ and width $\Delta U$. If $N_{ij}$ is the number of velocity pairs with $X$ in bin $i$ and $Y$ in bin $j$, the empirical joint and marginal probability masses are
\begin{equation}
    P_{ij}=\frac{N_{ij}}{N_xN_z},
    \qquad
    P_i=\frac{N_i}{N_xN_z},
    \label{eq:empirical_probabilities}
\end{equation}
where $N_i=\sum_jN_{ij}$.

At each bin centre, define the regularization window
\begin{equation}
    B_i=[c_i-\delta,c_i+\delta],
    \qquad
    \delta=\kappa\,\sigma_{u,\mathrm{plane}}(y,t_s),
    \qquad
    \kappa=0.1,
    \label{eq:regularisation_window}
\end{equation}
where
\begin{equation}
    \sigma_{u,\mathrm{plane}}(y,t_s)
    =
    \left\langle
    \left[
    u(x,y,z,t_s)-
    \langle u(x,y,z,t_s)\rangle_{xz}
    \right]^2
    \right\rangle_{xz}^{1/2}.
\end{equation}

This quantity is the instantaneous spatial standard deviation of the streamwise fluctuation within each wall-parallel plane, distinct from the ensemble r.m.s. profile $u'(y)$ used in the classical-structure thresholds. The estimator's regularization parameter $\kappa=0.1$ sets the window half-width to $10\,\%$ of this standard deviation, evaluated separately at each snapshot and wall distance.
 This adapts the smoothing to the local velocity variability rather than imposing a fixed width. The sensitivity to $\kappa$ is assessed in \S\ref{sec:parameter_choice} and Appendix~\ref{app:estimator_validation}. The overlap of the window $B_i$ with histogram bin $I_j$ is represented by

\begin{equation}
    W_{ij}=\frac{|B_i\cap I_j|}{\Delta U}.
    \label{eq:window_weights}
\end{equation}

The weights describe fractional bin overlap and are not separately normalized. The resulting marginal and joint density estimates are

\begin{subequations}\label{eq:regularised_densities}
\begin{align}
    \widetilde f_X(c_i)
    &=\frac{1}{2\delta}\sum_j W_{ij}P_j,\\
    \widetilde f_{XY}(c_i,c_k)
    &=\frac{1}{(2\delta)^2}
    \sum_{j,l}W_{ij}P_{jl}W_{kl}.
\end{align}
\end{subequations}

The conditional-density estimator is
\begin{equation}
    \widetilde f_{Y|X}(c_k\mid c_i)
    =
    \frac{\widetilde f_{XY}(c_i,c_k)+\varepsilon}
    {\widetilde f_X(c_i)+\varepsilon},
    \qquad
    \varepsilon=10^{-30}.
    \label{eq:conditional_density}
\end{equation}
Here, $\varepsilon$ is a numerical safeguard. Windows extending beyond the sampled velocity range are truncated without renormalization. The effect of truncation in $X$ cancels between the joint and marginal densities in the conditional-density ratio, whereas truncation in $Y$ can slightly underestimate the conditional density and hence overestimate surprisal near the range boundaries. The sensitivity to boundary truncation, histogram resolution and regularization is assessed in Appendix~\ref{app:estimator_validation}.

For an observed velocity pair belonging to bins $(i,k)$, the local conditional surprisal is
\begin{equation}
    S(\boldsymbol{x},t_s)
    =
    -\log\widetilde f_{Y|X}(c_k\mid c_i).
    \label{eq:surprisal_field}
\end{equation}
The natural logarithm is used, so that $S$ is expressed in nats. Because the estimate is evaluated at histogram-bin centres, $S$ is piecewise constant in velocity rather than obtained from a continuously interpolated density. As a differential conditional surprisal, its value depends on the velocity units and may be negative when the estimated density exceeds unity. These unit-dependent offsets are removed by the wall-normal standardization of \S\ref{sec:statistics}.

A large value of $S$ indicates that the observed velocity transition has a low conditional density given its initial state. Surprisal thus quantifies how statistically unusual that transition is in the prescribed convective frame, independently of the neural network's prediction error. However, it does not measure the intrinsic unpredictability of the complete flow.

The corresponding convected velocity increment is
\begin{equation}
    \Delta U_\ell
    =
    Y-X
    =
    U(x+\ell(y),y,z,t_{s+1})
    -
    U(x,y,z,t_s).
    \label{eq:convected_increment}
\end{equation}

Although surprisal is closely related to the convected velocity increment, it also accounts for the initial velocity from which the change occurs. Two transitions with the same increment can therefore have different surprisal values if their initial states differ and the corresponding changes have different conditional likelihoods. This distinction is examined quantitatively in \S\ref{sec:conditional_selection} and Appendix~\ref{app:increment}.

\subsection{Wall-normal standardization and structural statistics}
\label{sec:statistics}

The statistical properties of turbulent velocity fluctuations vary considerably with wall distance. To compare the relative unexpectedness of velocity transitions across the channel, we introduce the standardized conditional surprisal
\begin{equation}
    \widetilde{S}(\boldsymbol{x},t_s)
    =
    \frac{S(\boldsymbol{x},t_s)-\overline S(y)}
    {\sigma_S(y)},
    \label{eq:standardised_surprisal}
\end{equation}
where $\overline S(y)$ and $\sigma_S(y)$ are the mean and standard deviation of surprisal over the homogeneous directions and the available DNS snapshot-pair ensemble, evaluated independently at each wall-normal plane and Reynolds number. The two channel halves are not combined when constructing these reference profiles.

With this definition, $\widetilde{S}=0$ corresponds to the mean surprisal at that height, while $\widetilde{S}=1$ represents one local standard deviation above that mean. Positive values identify above-average surprisal and negative values below-average surprisal. The transformation preserves the ordering of surprisal values within each plane.

Fluctuation energy is expressed as the ratio $\langle E\rangle_A(y)/\langle E\rangle_{\mathrm{plane}}(y)$, providing an interpretable measure of intensity relative to the plane mean. For differential conditional surprisal, however, such a ratio would be inappropriate: surprisal can be negative, and changing the velocity units introduces an additive offset that alters the ratio to its mean. We therefore standardize surprisal using its wall-normal mean and standard deviation, removing this unit-dependent offset and expressing departures relative to the local distribution.

Wall-normal population statistics are obtained by pooling the relevant voxel sums and counts over the snapshot ensemble and, where appropriate, over the two channel halves before forming ratios. Let $\Pi(y)$ denote the wall-parallel planes at the same distance from their nearest wall, with the centreline counted only once, and let $\mathcal P_p$ denote the set of voxels in plane $p$. For a selected population $A(t)$, its mean standardized surprisal is
\begin{equation}
    \langle \widetilde{S}\rangle_A(y)
    =
    \frac{
    \displaystyle
    \sum_t\sum_{p\in\Pi(y)}
    \sum_{\boldsymbol{x}\in A(t)\cap\mathcal P_p}
    \widetilde{S}(\boldsymbol{x},t)
    }{
    \displaystyle
    \sum_t\sum_{p\in\Pi(y)}
    |A(t)\cap\mathcal P_p|
    },
    \label{eq:mean_surprisal_region}
\end{equation}
provided that the denominator is non-zero. Pooling voxel sums and counts before forming the mean gives equal weight to every voxel in the population, so snapshots contribute in proportion to the number of voxels they contain in that population. This characterizes the selected fluid as a whole, rather than averaging instantaneous population means with equal weight per snapshot. Population energy means are defined analogously, with $E$ replacing $\widetilde{S}$. Aggregate contributions to plane-wide energy, attribution magnitude and surprisal totals are defined in Appendix~\ref{app:contributions}.

Uncertainties are estimated using a temporal block bootstrap \citep{Kunsch1989}, with blocks of 120 consecutive stored DNS fields. This length exceeds twice the estimated integral correlation time for the principal population contrasts, selection factors and interface contrasts at both Reynolds numbers. Confidence intervals are obtained by resampling these blocks and recomputing the statistics from accumulated voxel counts and sums. Comparisons between populations use paired bootstrap resamples. The sensitivity to block length is examined in Appendix~\ref{app:bootstrap}.

\subsection{Choice of estimator parameters}
\label{sec:parameter_choice}

The estimator depends on the number of velocity bins, $N_b$, and on the regularization parameter $\kappa$. We adopt $N_b=120$ and $\kappa=0.1$ at both Reynolds numbers (Appendix~\ref{app:estimator_validation}). In a Gaussian benchmark with known conditional entropy, this configuration yields a mean absolute error of $4.9\times10^{-3}$ nats and a maximum absolute error of $7.3\times10^{-3}$ nats across the tested cases. A wider window with more bins is marginally more accurate in that benchmark, but introduces greater smoothing in velocity space. In the DNS, surprisal fields remain highly correlated across the tested resolutions. As expected for an extreme-tail comparison, the identity of the uppermost-surprisal voxels is more sensitive: their median Jaccard overlap with a $400$-bin reference is approximately $0.86$--$0.90$. The conditioned analyses therefore use the median surprisal of each plane and snapshot rather than an extreme-percentile threshold. The principal SHAP-only selection factor remains stable under the tested variations of $N_b$ and $\kappa$, supporting the adopted parameter choice (Appendix~\ref{app:dns_validation}).

\subsection{High- and low-speed pairs within SHAP structures}
\label{sec:pairs}

The analyses of \S\S\ref{sec:shap_signatures}--\ref{sec:conditional_selection} concern individual voxels, irrespective of the connected structure to which they belong. We now examine complete connected components of the SHAP mask $\mathcal H$, focusing on their high- and low-speed content and the properties of their internal interfaces. The classification, spatial arrangement and principal interfacial signatures are examined at both Reynolds numbers, while selected geometrical and kinematic controls are evaluated on smaller subsets.

We retain connected SHAP structures with volume $V^+>30^3$, following \citet{MolinaCasino2026}, and classify them by the volume-fraction procedure introduced in that study, using a purity threshold of $80\,\%$. For each structure, the nearest wall is identified from the position of its volume-weighted centroid. The classification criteria are applied sequentially. Structures in which more than $80\,\%$ of their volume satisfies $|uv|>1.62\,u'(y)v'(y)$, considering all four quadrants together, are assigned to the intense quadrant-event family. Each structure is labelled according to the quadrant containing the largest volume satisfying this intensity criterion, with no minimum fraction required for the dominant quadrant. All volume fractions are computed using physical cell volumes. Otherwise, a structure is classified as high-speed or low-speed when the corresponding sign of $u$ occupies more than $80\,\%$ of its volume. The remaining structures are classified as updrafts or downdrafts when more than $80\,\%$ of the volume has the corresponding sign of the wall-oriented $v$ component. Structures satisfying none of these conditions form the residual class of \emph{high-/low-speed pairs}. The threshold $1.62$ is specific to this classification and differs from $\beta=1.75$, used to define the classical Q-event mask. For structures referred to the upper wall, the sign of $v$ is reversed consistently during classification.

Within each pair, high-speed (HS) and low-speed (LS) fluid are defined by $u>0$ and $u<0$, respectively. Their volume-weighted centroids account for periodicity in the homogeneous directions. The displacement $\boldsymbol d=(d_x,d_y,d_z)$ is measured from the LS centroid to the HS centroid, with the wall-normal direction referred to the nearest wall. Pairs with centroid separation $|\boldsymbol d|^+\leq5$ or $|\boldsymbol d|^+\geq400$ are left untyped. These untyped pairs are omitted from the orientation-specific statistics. A separate sensitivity check shows that including them has a negligible effect on the aggregate interface-minus-lobe contrast.

\begin{figure}
    \centering
    \includegraphics[width=0.85\textwidth]{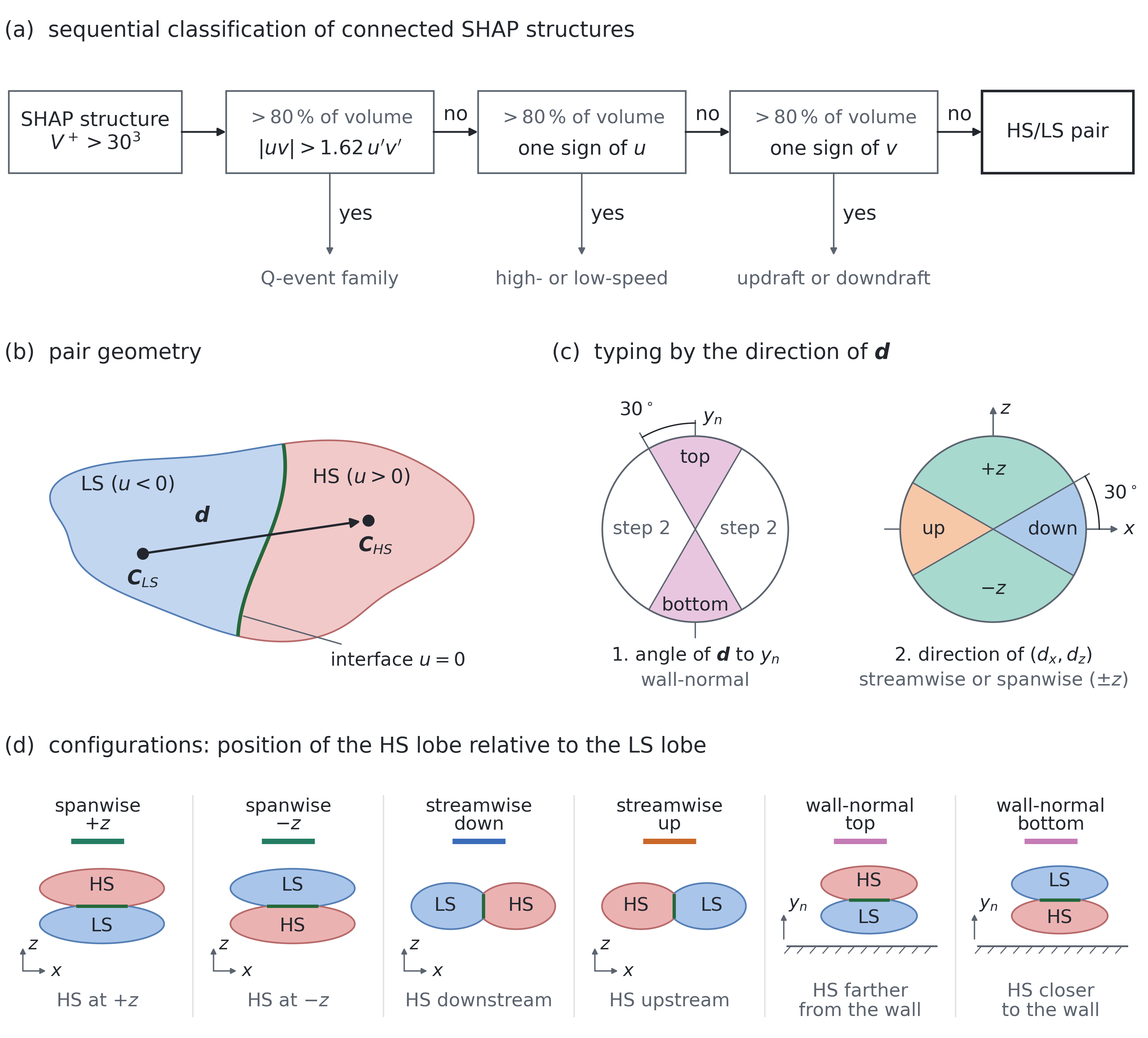}
    \caption{Schematic of the classification of connected SHAP structures and high-/low-speed (HS/LS) pairs. (a) Sequential classification according to the dominant velocity characteristics, with HS/LS pairs forming the residual class. (b) Definition of the HS and LS regions, their centroids and the displacement vector $\boldsymbol d$ from LS to HS, together with the internal $u=0$ interface. (c) Orientation classification based on the wall-normal direction and the projection of $\boldsymbol d$ onto the $(x,z)$ plane. (d) The six configurations defined by the position of the HS region relative to the LS region. Pairs with centroid separations outside $5<|\boldsymbol d|^+<400$ are left untyped. All geometries are schematic.}
    \label{fig:pair_classification}
\end{figure}

The remaining pairs are classified according to the direction of $\boldsymbol d$. Pairs within $30^\circ$ of the wall-normal direction are classified as wall-normal. Otherwise, pairs whose projection onto the $(x,z)$ plane lies within $30^\circ$ of the streamwise axis are classified as streamwise, and all remaining pairs as spanwise. 
These orientation classes cover unequal solid angles. For isotropically distributed centroid displacements, the corresponding geometric fractions are approximately $13.4\,\%$, $28.9\,\%$ and $57.7\,\%$ for wall-normal, streamwise and spanwise orientations, respectively. The six configurations are defined by the position of the HS lobe relative to the LS lobe: \emph{streamwise-up} and \emph{streamwise-down} indicate that HS lies upstream or downstream, respectively; \emph{spanwise $+z$} and \emph{spanwise $-z$} indicate its displacement in the positive or negative spanwise direction; and \emph{wall-normal-top} and \emph{wall-normal-bottom} indicate that HS lies farther from or closer to the nearest wall, respectively (figure~\ref{fig:pair_classification}).

For each pair, the principal HS and LS lobes are identified as the largest six-connected components, by volume, of the positive- and negative-$u$ regions. Their internal interface is defined by contacts between these lobes, with the $u=0$ crossing located by linear interpolation between adjacent voxels of opposite sign. Only pairs with contacting principal lobes contribute to the interfacial analyses. The interface is distinct from the outer boundary of the SHAP structure and excludes crossings associated exclusively with smaller secondary lobes.

The interface centre $\boldsymbol C_I$ is defined by the area-weighted mean of its crossing points. A characteristic normal $\hat{\boldsymbol n}$ is defined as the eigenvector associated with the smallest eigenvalue of the area-weighted covariance matrix of the interface crossing points, oriented from the LS towards the HS lobe. Principal-lobe interfaces and their distances are evaluated without periodic wrapping; their reconstruction and geometrical sensitivity are assessed in Appendices~\ref{app:interface_distance} and \ref{app:interface_composition}.

To examine surprisal across the interface, we assign each voxel of a pair a signed distance $d_{u0}^+$, defined as the minimum three-dimensional Euclidean distance, in wall units, to the interpolated crossing points of the principal-lobe interface. Distances are negative in the LS region and positive in the HS region. Profiles of standardized surprisal $\widetilde{S}$ are accumulated in bins of 2 wall units and pooled over the analysed pairs with volume weighting, accounting for the non-uniform wall-normal grid. We define the interfacial region by $|d_{u0}^+|\leq2$ and the lobe-interior reference by $20\leq|d_{u0}^+|\leq45$. The interface-minus-lobe contrast is defined as the difference between the mean standardized surprisal in the interfacial region and the larger of the two lobe-interior means. The aggregate statistics include all six pair arrangements.

To distinguish alignment with the interface from alignment with the structure as a whole, we also construct internally standardized profiles for each pair. Distances are binned within $\pm200$ wall units, using bins of $8$ wall units from the interface and one streamwise grid spacing from the volume-weighted streamwise centroid of the complete SHAP structure. For pair $i$, bins containing at least $20$ voxels are retained, and pairs with fewer than three resolved bins or zero profile variance are excluded. The internal standardized profile is
\begin{equation}
    Z_{i,k}=\frac{\langle \widetilde{S}\rangle_{i,k}-\mu_i}{\sigma_i},
    \label{eq:internal_z}
\end{equation}
where $\langle \widetilde{S}\rangle_{i,k}$ is the volume-weighted mean surprisal in bin $k$, and $\mu_i$ and $\sigma_i$ are the unweighted mean and sample standard deviation across the resolved bins of pair $i$. The resulting profiles are averaged with equal weight per pair. To test whether the observed peak is specifically aligned with the HS/LS interface, rather than arising from the averaging of unrelated profile fluctuations, we construct a geometric null by independently circularly shifting each pair's profile before averaging. At $Re_\tau\simeq125$, confidence intervals for the original profiles are estimated from $400$ bootstrap realizations, resampling pairs in $67$ temporal blocks spanning $120$ consecutive fields of the original DNS record. The geometric null is evaluated using another $400$ realizations with randomly shifted profiles and the same block-resampling procedure.

Surprisal for the transition from $t_s$ to $t_{s+1}$ is assigned to the initial instant $t_s$, consistently with the SHAP structures at that instant; alternative assignments are examined in Appendix~\ref{app:temporal}, the distance construction in Appendix~\ref{app:interface_distance}, and the role of the convected increment and of generic $u=0$ surfaces in Appendix~\ref{app:increment}. For wall-normal comparisons, pairs are grouped by the wall distance of their volume-weighted centroid into $y^+<20$, $20$--$40$, $40$--$80$ and $y^+\geq80$.

Finally, to characterize the physical composition of the internal interface, the $u=0$ surface between the principal HS and LS lobes is reconstructed as a triangulated isosurface \citep{Lewiner2003}. Its area is classified according to the SHAP and classical coherent-structure masks defined in \S\ref{sec:structures}. Reconstruction details and sensitivity checks are provided in Appendix~\ref{app:interface_composition}.

\section{Results}
\label{sec:results}

We first examine the SHAP-selected volume voxel by voxel, comparing its energetic and informational signatures with those of classical coherent structures. We then test whether conditional surprisal distinguishes selection at comparable energy and wall distance and characterize the velocity changes associated with the selected non-classical fluid. The subsequent sections address complete SHAP structures and their internal high-/low-speed interfaces.

\subsection{Selected fluid inside and outside the classical masks}
\label{sec:shap_signatures}

\begin{figure}
    \centering
    \includegraphics[width=\textwidth]{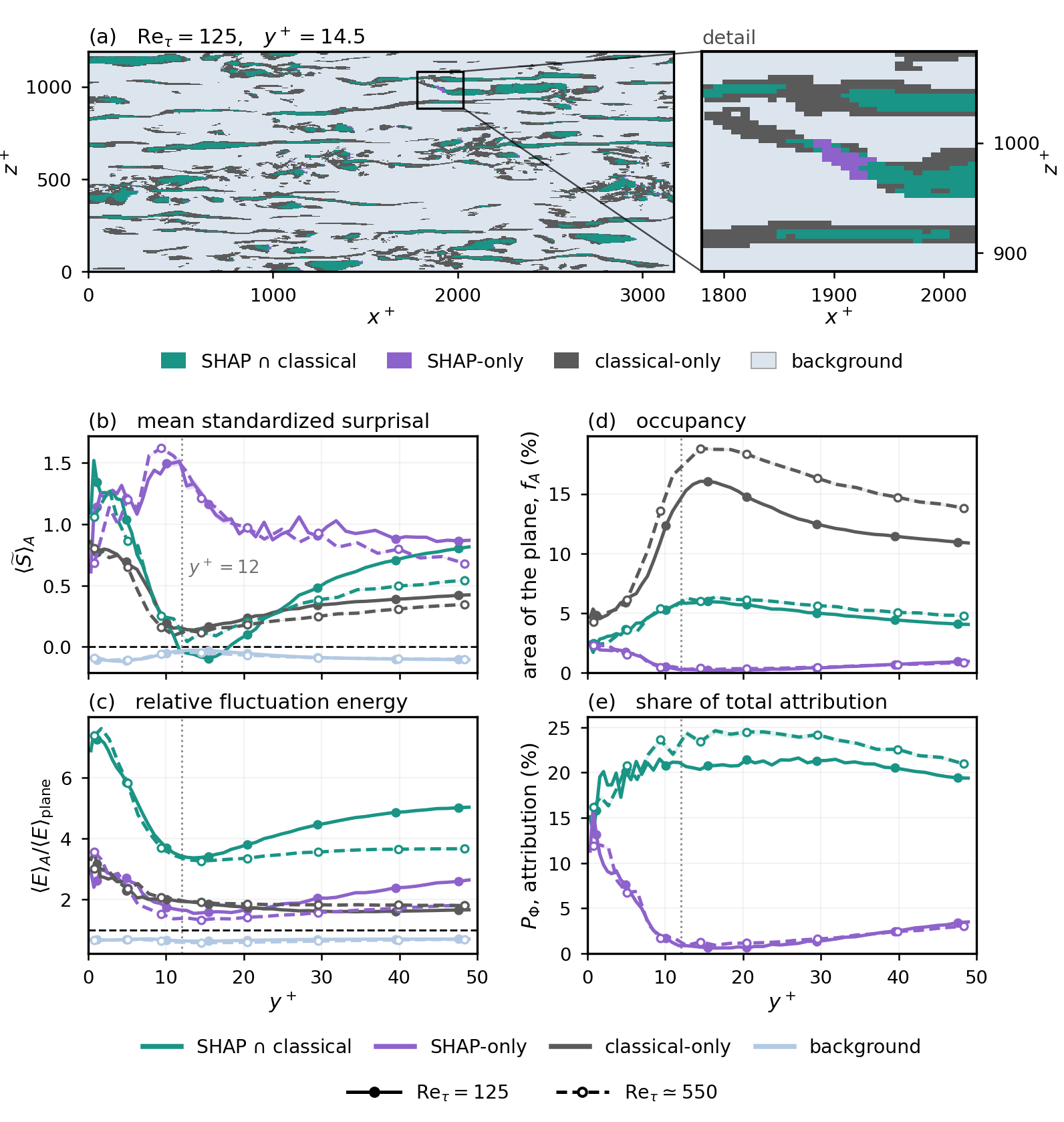}
    \caption{Energetic and informational signatures of SHAP-selected fluid at $Re_\tau \simeq 125$ (solid lines, filled symbols) and $Re_\tau \simeq 550$ (dashed lines, open symbols). (a) A wall-parallel section at $Re_\tau\simeq125$ and $y^+\simeq14.5$, showing the four mutually exclusive populations, with an enlarged view of the selected regions. (b) Population mean standardized surprisal $\langle \widetilde{S}\rangle_A$ and (c) relative fluctuation energy $\langle E\rangle_A/\langle E\rangle_{\mathrm{plane}}$. (d) Voxel occupancy $f_A$, excluding background (approximately $75$--$91\,\%$ of the plane) for clarity. (e) Fraction $P_\Phi$ of the total plane-wide attribution magnitude contributed by the two SHAP-selected populations (Appendix~\ref{app:contributions}). The vertical reference marks $y^+=12$, below which the prescribed convection velocity is constant. Shaded regions indicate $95\,\%$ block-bootstrap confidence intervals.}
    \label{fig:shap_energy_surprisal}
\end{figure}

\citet{Cremades2025} showed that regions identified as important for neural-network predictions overlap only partially with classical coherent structures, leaving a portion of the selected fluid outside conventional criteria. In addition, the authors reported limited agreement with vortical structures, without a clear explanation of the physical mechanism. What distinguishes this fluid, referred to here as SHAP-only, remains unclear. To investigate its statistical properties, we compare the four mutually exclusive populations defined in \S\ref{sec:structures} in terms of fluctuation energy and mean standardized surprisal. Figure~\ref{fig:shap_energy_surprisal}(a) shows their distribution in wall-parallel sections at $y^+\simeq14.5$. SHAP-only voxels occupy relatively small, scattered patches, often adjacent to fluid selected by both SHAP and the classical criteria. These observations concern individual voxels and do not imply that SHAP-only voxels form independent connected structures.

The clearest distinction emerges between the two SHAP-selected populations [figure~\ref{fig:shap_energy_surprisal}(b,c)]. At $Re_\tau\simeq125$, the SHAP-classical overlap is strongly energetic, containing approximately $3.4$--$7$ times the plane-mean fluctuation energy. SHAP-only regions are less energetic, with approximately $1.5$--$3$ times the plane mean, but are far from quiescent. Remarkably, this energetic ordering is reversed for surprisal. The mean standardized surprisal of SHAP-only voxels is positive throughout the heights considered and reaches approximately $1.5$ near $y^+\simeq12$, whereas that of the classical overlap approaches zero and becomes slightly negative around $12\lesssim y^+\lesssim17$. Except immediately next to the wall ($y^+\lesssim3$), SHAP-only exhibits the highest mean surprisal of the four populations.

This contrast reveals a fundamental difference in the observed evolution of the selected fluid. The SHAP-classical overlap is characterized by particularly high fluctuation energy, whereas SHAP-only voxels, despite being less energetic, undergo substantially more unusual velocity transitions relative to the statistics at the same wall distance. Thus, the selected fluid does not have a uniform energetic or informational signature: instantaneous fluctuation intensity alone does not fully describe the velocity transitions occurring in all the regions highlighted by the forecasting network.

The predominance of classical fluid within the complete SHAP mask partly conceals this distinction. At $Re_\tau\simeq125$, more than $90\,\%$ of the SHAP-selected volume between $y^+\simeq10$ and $33$ belongs to at least one classical mask, with an overlap of approximately $96\,\%$ between $y^+\simeq12$ and $22$ [figure~\ref{fig:shap_energy_surprisal}(d)]. This extensive overlap is largely associated with the presence of streaks in the near-wall region. As a result, the mean surprisal of the complete SHAP mask remains close to the plane average in the buffer layer, concealing the higher surprisal of the smaller SHAP-only population.

A population's mean surprisal must also be distinguished from its contribution to the total selected volume and attribution. Between $y^+=12$ and $20$, SHAP-only occupies approximately $0.2\,\%$ of the plane and accounts for only $0.6$--$0.9\,\%$ of its total attribution magnitude, compared with approximately $20$--$21\,\%$ for the SHAP-classical overlap [figure~\ref{fig:shap_energy_surprisal}(d,e)]. Classical streaks likewise contribute substantially larger shares of energy and attribution, despite having mean standardized surprisal below the plane reference (Appendix~\ref{app:contributions}). SHAP-only is therefore distinguished by its high mean surprisal, rather than by its overall volume or contribution to the attribution budget.

The same qualitative distinction persists at $Re_\tau\simeq550$. SHAP-only again exhibits the highest mean surprisal away from the immediate vicinity of the wall, reaching approximately $1.6$ near $y^+\simeq9$, where turbulent motions become increasingly prominent. Its relative fluctuation energy is $1.3$--$3.6$ times the plane mean, compared with $3.3$--$7.6$ for the SHAP-classical overlap. Between $y^+\simeq9$ and $33$, the classical overlap still accounts for approximately $91.6$--$95.9\,\%$ of the SHAP-selected volume. Unlike at the lower Reynolds number, the complete SHAP mask has moderately positive mean surprisal in the buffer layer. Nevertheless, the main distinction remains: classical-overlap fluid dominates the selected volume and is more energetic, whereas SHAP-only voxels exhibit a stronger surprisal signature.


\subsection{Surprisal and selection outside the classical masks}
\label{sec:conditional_selection}

We now examine whether surprisal provides information about SHAP selection beyond its strong association with fluctuation energy. To separate the two effects statistically, we make two complementary comparisons: first, we compare selection probabilities at similar energy and wall distance but different surprisal; then we reverse the comparison, fixing surprisal and wall distance while varying energy.

For the first comparison, energy percentiles are computed separately for each wall-parallel plane and snapshot and grouped into eight strata, indexed by $m=1,\ldots,8$ in increasing order
of energy: 0--50, 50--70, 70--80, 80--90, 90--92, 92--96, 96--98 and 98--100. The finer strata at high energy allow us to compare voxels with more similar energy levels in the range where the fraction selected by SHAP increases most sharply. Within each stratum, voxels are separated according to whether $\widetilde{S}$ is above ($S^>$) or below ($S^<$) the median of the complete plane and snapshot. The median is not recalculated within individual energy strata. We define the surprisal selection factor as
\begin{equation}
    R_{S\mid E}(A;k,m)=
    \frac{P(A\mid S^>,k,m,\mathcal U_A)}
         {P(A\mid S^<,k,m,\mathcal U_A)},
    \label{eq:selection_ratio}
\end{equation}
where $k$ denotes a wall-normal band and $m$ an energy stratum. The reference universe $\mathcal U_A$ specifies which voxels are eligible for selection: the entire flow for the complete SHAP mask, only voxels satisfying at least one classical criterion for the SHAP-classical overlap, and only voxels outside all classical masks for SHAP-only. Within each universe, selection probability is the number of SHAP-selected eligible voxels divided by the total number of eligible voxels in the corresponding surprisal, energy and height group. Thus, $R_{S\mid E}>1$ indicates that above-median surprisal is associated with more frequent SHAP selection at comparable energy and wall distance. To obtain an overall factor while accounting for differences in energy, we use the Mantel--Haenszel risk ratio \citep{MantelHaenszel1959,GreenlandRobins1985},

\begin{equation}
    R_{S\mid E}(A;k)=
    \frac{\displaystyle\sum_m a_{1m}r_{0m}/N_m}
         {\displaystyle\sum_m a_{0m}r_{1m}/N_m},
    \label{eq:mh_selection}
\end{equation}

where $a_{1m}$ and $a_{0m}$ are the numbers of selected voxels with above- and below-median surprisal in stratum $m$, $r_{1m}$ and $r_{0m}$ are the corresponding numbers of eligible voxels, and $N_m=r_{1m}+r_{0m}$. Strata containing less than $1\,\%$ of the selected population's volume are excluded. In $5<y^+<12$, seven of the eight energy strata are retained for both SHAP-only and classical-overlap comparisons at each Reynolds number. These strata contain at least $99.79\,\%$ of the selected voxels in each population, while covering $91.9\,\%$ and $93.4\,\%$ of the eligible classical-overlap voxels at $Re_\tau\simeq125$ and $550$, respectively.

For the complementary comparison, voxels are grouped into ten equal-frequency surprisal strata (deciles), again defined separately for each plane and snapshot. Within each decile, we compare the selection probabilities of voxels with fluctuation energy above and below the plane median. This gives the energy selection factor $R_{E\mid S}$. The two calculations determine how strongly each descriptor remains associated with selection after controlling for the other, without assuming that the neural network implements successive selection criteria.

\begin{figure}
    \centering
    \includegraphics[width=0.8\textwidth]{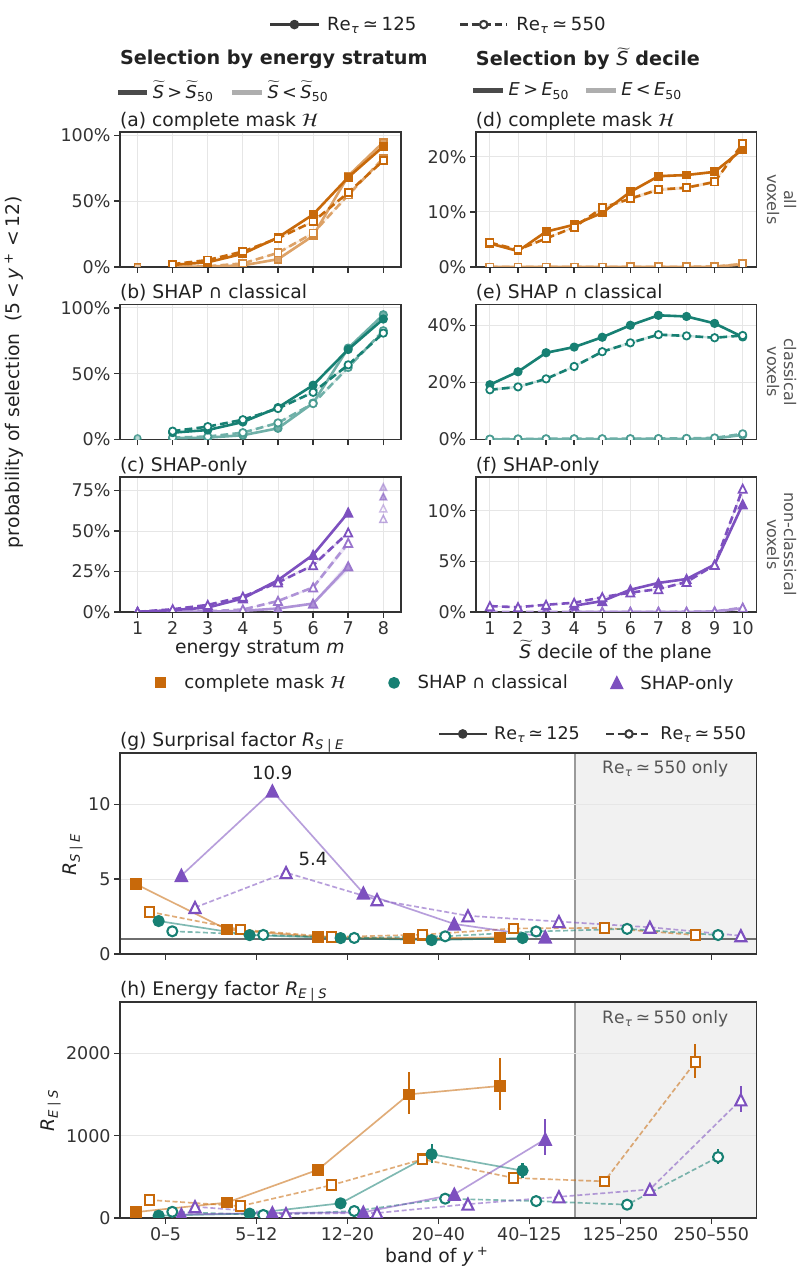}
    \caption{Conditional SHAP selection probabilities and their associations with fluctuation energy and standardized surprisal at $Re_\tau\simeq125$ (solid lines, filled symbols) and $Re_\tau\simeq550$ (dashed lines, open symbols). (a--c) Selection probabilities across energy strata for $5<y^+<12$, comparing voxels with surprisal above (dark) and below (light) the plane median. (d--f) Selection probabilities across surprisal deciles in the same wall-normal band, comparing voxels with fluctuation energy above (dark) and below (light) the plane median. (g) Surprisal factor $R_{S\mid E}$ and (h) energy factor $R_{E\mid S}$ as functions of wall-normal distance, with error bars indicating $95\,\%$ confidence intervals from the temporal block bootstrap described in \S\ref{sec:statistics}.}
    \label{fig:selection_ratio}
\end{figure}

The first result is the dominant association between fluctuation energy and SHAP selection. Across all three eligible populations, selection probabilities rise sharply with energy [figure~\ref{fig:selection_ratio}(a--c)]. The reverse comparison confirms this behaviour: even at comparable surprisal, high-energy voxels are selected much more frequently than low-energy voxels, often by orders of magnitude [figure~\ref{fig:selection_ratio}(d--f,h)]. Thus, the overall preference of SHAP for energetic fluid is not explained by differences in surprisal.

The second result reveals an additional distinction. Within comparable energy strata, voxels with above-median surprisal are generally selected more frequently than those with below-median surprisal [figure~\ref{fig:selection_ratio}(a--c)]. This association is especially pronounced outside the classical masks. In $5<y^+<12$, the pooled surprisal factor defined in \eqref{eq:mh_selection} is approximately $10.9$ for SHAP-only at $Re_\tau\simeq125$ and $5.4$ at $Re_\tau\simeq550$, compared with only $1.26$ and $1.27$, respectively, within the classical overlap [figure~\ref{fig:selection_ratio}(g)]. Although the precise values depend on the structure-identification thresholds, the stronger near-wall association between surprisal and SHAP-only selection persists under the variations examined in Appendix~\ref{app:threshold_sensitivity}. At $Re_\tau\simeq125$, the SHAP-only selection factor decreases to approximately $4.1$ in $12<y^+<20$ and $2.0$ in $20<y^+<40$. The association with surprisal weakens with increasing wall distance at both Reynolds numbers, although the decrease is more gradual at $Re_\tau\simeq550$. The weaker association near the channel centre is more sensitive to histogram resolution (Appendix~\ref{app:estimator_validation}). The near-wall association is also robust to finer matching: using $200$ snapshots per Reynolds number, increasing the number of energy strata from eight to fifty and comparing voxels within individual wall-normal planes changes the SHAP-only selection factors by less than $6\,\%$ across all bands, and by less than $2\,\%$ in the key region $5<y^+<12$.

The association between surprisal and SHAP-only selection is closely related to the convected velocity increment. At $Re_\tau\simeq125$, surprisal and increment magnitude both discriminate selected from unselected voxels outside the classical masks at comparable energy and wall distance, with AUC values of $0.88$ and $0.85$, respectively. When increment magnitude is also controlled, the AUC for surprisal decreases to $0.63$, indicating that much of its discrimination is shared with the increment (Appendix~\ref{app:increment}). A separate control using $200$ snapshots per Reynolds number examines the signed increment. In $5<y^+<12$, the selection factors remain $7.1$ and $2.7$ at the two Reynolds numbers. On the same retained cells, however, the energy-only factors are already $6.7$ and $2.5$, showing that the decrease relative to the original analysis is mainly due to the reduced common support. Thus, surprisal remains strongly associated with near-wall SHAP-only selection within the comparable populations, although its variation is closely related to the convected increment.

At matched energy, SHAP-only selection is also associated with smaller $|u|$ and larger $|w|$ (Appendix~\ref{app:increment}). The relevance of spanwise motions to wall turbulence is illustrated by the response to imposed spanwise wall oscillations reported by \citet{Marusic2021}, although that study concerns flow control rather than naturally occurring fluctuations.

Together, these results reveal two complementary signatures of SHAP selection. Fluctuation energy dominates across the complete flow, consistent with the extensive overlap between SHAP and classical intensity-based structures. Yet SHAP-only voxels, although less energetic than the classical overlap, exhibit a strong association between surprisal and selection even at comparable energy and wall distance. This association is largely related to the convected velocity increment, while retaining some discrimination beyond increment magnitude alone.

These selection factors measure conditional associations, not successive decisions made by the network. A large factor does not necessarily imply a high absolute selection probability: low-energy voxels outside the classical masks remain relatively unlikely to be selected, even when their surprisal is elevated [figure~\ref{fig:selection_ratio}(c)]. Furthermore, $R_{S\mid E}$ and $R_{E\mid S}$ represent different conditional comparisons and cannot be combined to quantify independent fractions of predictive importance.

Two additional controls support the robustness of this association. It persists under other observation frames, although its magnitude is reduced for a fixed observer (Appendix~\ref{app:robustness}). It is also not explained solely by proximity to classical structures: matching wall distance, energy, distance to the classical masks and local velocity-gradient magnitude still yields a SHAP-only selection factor of $8.8$ (Appendix~\ref{app:margins}).

\subsection{High- and low-speed pairs within SHAP-selected structures}
\label{sec:shap_pairs}

The preceding voxel-wise analysis shows that SHAP selection is strongly associated with fluctuation energy, while conditional surprisal provides additional statistical discrimination, particularly outside the classical coherent-structure masks. We now turn to the organization of complete SHAP-selected structures. In an adverse-pressure-gradient boundary layer over a wing, \citet{MolinaCasino2026} identified connected SHAP structures containing adjacent high- and low-speed fluid, with attribution concentrated near their internal boundary. We apply their classification to turbulent channel flow at both Reynolds numbers to characterize the occurrence, orientation and wall-normal distribution of comparable structures. Unlike the preceding analysis, this classification concerns complete SHAP structures, which may contain both classical-overlap and SHAP-only regions.

High-/low-speed (HS/LS) pairs account for $8.3\,\%$ of the retained SHAP structures and $11.9\,\%$ of their combined volume at $Re_\tau\simeq125$. Similar proportions are found at $Re_\tau\simeq550$, with $9.4\,\%$ of structures and $11.7\,\%$ of the volume. More than $99\,\%$ of the pairs can be assigned to one of the six configurations, and over $92\,\%$ of the classified pairs contribute to the interfacial profiles.

\begin{figure}
    \centering
    \includegraphics[width=0.95\textwidth]{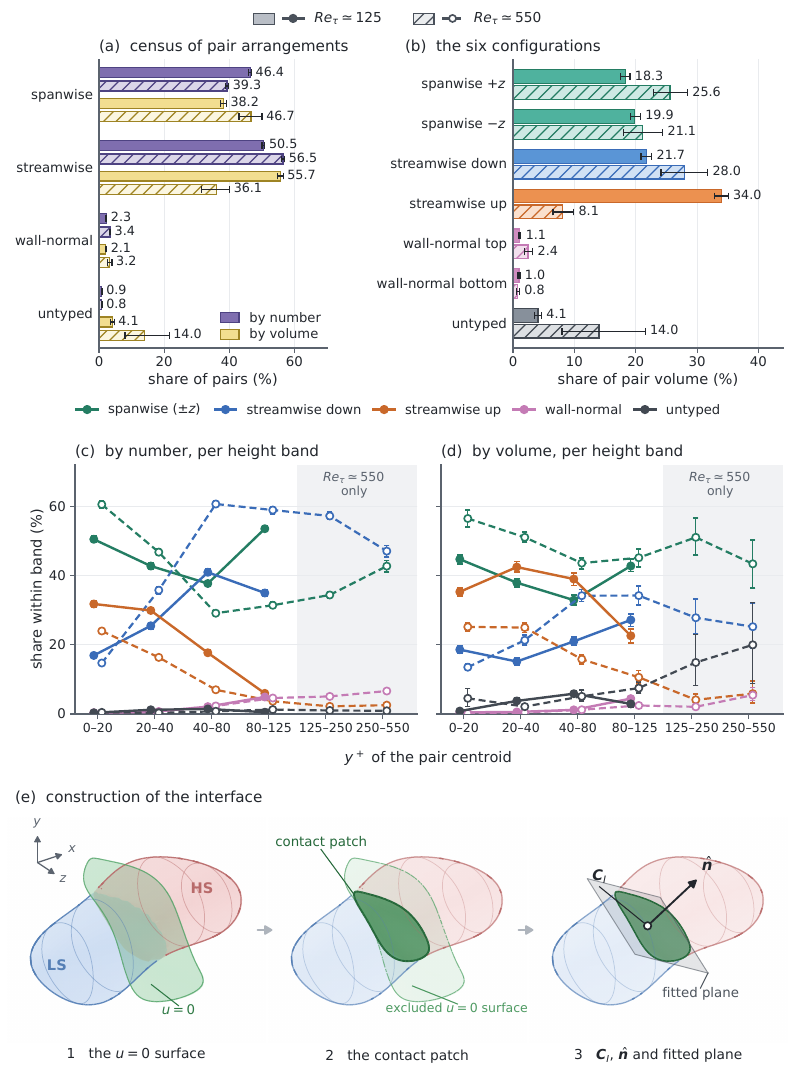}
    \caption{Census and spatial organization of SHAP-selected high-/low-speed (HS/LS) pairs at $Re_\tau\simeq125$ and $550$. (a) Fractions of pairs by number and combined volume for the principal orientations, including untyped pairs. (b) Volume fractions of the six resolved configurations and untyped pairs; the geometrical arrangements are illustrated in figure~\ref{fig:pair_classification}(d). (c,d) Distribution of pair orientations by number and volume across wall-normal bands defined by the pair centroids. Bands above $y^+=125$ occur only at $Re_\tau\simeq550$. (e) Schematic construction of the principal-lobe $u=0$ interface, its contact patch and a fitted plane defining the interface centre $\boldsymbol C_I$ and characteristic normal $\hat{\boldsymbol n}$. The fitted plane characterizes the interface orientation rather than replacing its reconstructed geometry. Filled bars and symbols correspond to $Re_\tau\simeq125$, and hatched bars and open symbols to $Re_\tau\simeq550$. Error bars indicate $95\,\%$ block-bootstrap confidence intervals.}
    \label{fig:shap_pairs}
\end{figure}

At $Re_\tau\simeq125$, streamwise arrangements are the most frequent, representing $50.5\,\%$ of HS/LS pairs, followed by spanwise ($46.4\,\%$) and wall-normal ($2.3\,\%$) arrangements. Streamwise pairs also dominate by volume, contributing $55.7\,\%$ of the total pair volume, compared with $38.2\,\%$ for spanwise pairs [figure~\ref{fig:shap_pairs}(a)]. Relative to the isotropic reference described in \S\ref{sec:pairs}, streamwise arrangements are overrepresented and wall-normal arrangements underrepresented. These differences cannot therefore be attributed solely to the unequal angular ranges used to define the orientation classes.

The relative positions of the lobes reveal further differences [figure~\ref{fig:shap_pairs}(b--d)]. At $Re_\tau\simeq125$, streamwise-up pairs account for $34.0\,\%$ of the pair volume, compared with $21.7\,\%$ for streamwise-down pairs. This differs from the wing boundary layer examined by \citet{MolinaCasino2026}, where spanwise pairs occupy more volume and the opposite streamwise ordering predominates. At $Re_\tau\simeq550$, streamwise-down pairs account for a larger fraction of the total pair volume, although streamwise-up configurations remain more prevalent near the wall. The preferred arrangement therefore depends on wall distance and flow configuration, while adjacent HS and LS lobes are found in both channels.

The near-wall occurrence of streamwise-up pairs is geometrically reminiscent of the high-speed fluid overtaking low-speed fluid associated with causal significance by \citet{OsawaJimenez2024}. However, the present classification describes the relative positions of the lobes, not their temporal motion or causal influence. These pairs provide a basis for examining the velocity transitions near their internal boundaries.

\subsection{Surprisal at the pair interface}
\label{sec:pair_interfaces}

The internal boundary between the principal HS and LS lobes is particularly interesting because \citet{MolinaCasino2026} found that ensemble-averaged SHAP attribution was concentrated near the surface separating them in a wing boundary layer. At this interface, the streamwise velocity fluctuation vanishes, although transverse fluctuations, velocity gradients and temporal changes need not be negligible. We investigate the observed velocity transitions near these boundaries using conditional surprisal, which is computed directly from the DNS without relying on the forecasting network or its attributions.

\begin{figure}
    \centering
    \includegraphics[width=0.9\textwidth]{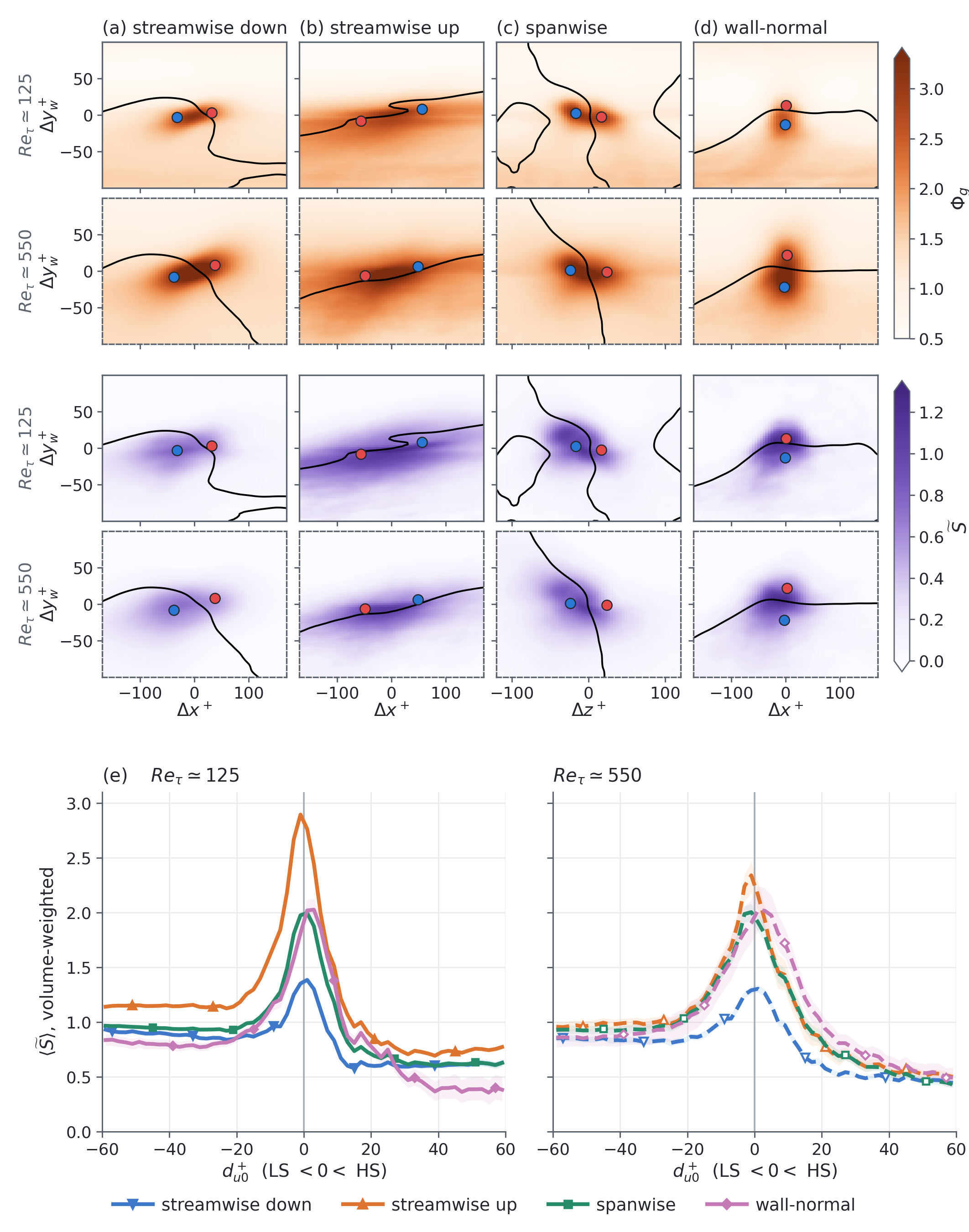}
    \caption{SHAP attribution and standardized conditional surprisal around HS/LS pairs at $Re_\tau\simeq125$ and $550$. (a--d) Equal-weight-per-pair composites for streamwise-down, streamwise-up, spanwise ($+z$) and wall-normal (HS above LS) configurations. Each pair contributes a plane through the midpoint of its HS and LS centroids: $x$--$y$ for streamwise and wall-normal pairs, and $z$--$y$ for spanwise pairs. Upper-wall pairs are reflected in $y$; opposite orientations are analysed separately. Pixels are averaged over pairs with valid data, excluding locations outside the channel. The upper two rows show normalized attribution magnitude $\Phi_g$, and the lower two show standardized surprisal $\widetilde{S}$, at both Reynolds numbers. Black contours indicate $u=0$ in the ensemble-mean velocity field; blue and red symbols mark the mean projected LS and HS centroids. Because fields and centroids are averaged separately, the symbols need not lie on opposite sides of the contours. Colour-bar extensions indicate values outside the displayed ranges. The numbers of pairs at $Re_\tau\simeq125$ are $36\,147$, $23\,320$, $26\,024$ and $1648$, and at $Re_\tau\simeq550$ are $71\,001$, $12\,421$, $28\,933$ and $3347$, in panel order. (e) Volume-weighted mean $\widetilde{S}$ against signed distance $d_{u0}^+$ from the HS/LS interface, negative on the LS side. Shading denotes $95\,\%$ temporal block-bootstrap confidence intervals. The numbers of pairs contributing within $|d_{u0}^+|\leq2$, across all six configurations, range from $955$ to $34\,053$ at $Re_\tau\simeq125$ and from $1583$ to $66\,749$ at $Re_\tau\simeq550$.}
    \label{fig:pair_interfaces}
\end{figure}

The ensemble averages in figure~\ref{fig:pair_interfaces}(a--d) show higher standardized surprisal near the boundary between the HS and LS lobes. This region extends farther in streamwise-up pairs, whereas the spanwise and wall-normal arrangements show more localized concentrations. The attribution composites show a related organization around the velocity-sign transition, although their maxima do not coincide everywhere with those of surprisal. Notably, a statistical measure derived from the observed velocity transitions highlights the same general interfacial region previously identified as important in SHAP-selected structures. Since ensemble averaging can obscure the geometry of individual pairs, we next quantify the enhancement using distances measured from each pair's own reconstructed interface.

The profiles in figure~\ref{fig:pair_interfaces}(e) reveal a pronounced maximum of standardized surprisal at the boundary between the HS and LS lobes. Here, $d_{u0}^+$ denotes the signed minimum three-dimensional distance, in wall units, to the interpolated $u=0$ crossings between the lobes, negative in the LS region and positive in the HS region (see \S\ref{sec:pairs}). Averaging over all six pair arrangements, the mean standardized surprisal at $Re_\tau\simeq125$ reaches $1.95$ within $|d_{u0}^+|\leq2$, compared with $0.97$ in the LS interior and $0.65$ in the HS interior, corresponding to an interface-to-LS difference of $0.98$. At $Re_\tau\simeq550$, the corresponding values are $1.65$, $0.90$ and $0.58$, giving a difference of $0.75$. Thus, the interfacial enhancement of standardized surprisal is observed at both Reynolds numbers. This enhancement persists when the interfacial region is widened to $|d_{u0}^+|\leq4$ and $8$, with interface-minus-lobe differences of $0.90$ and $0.67$ at $Re_\tau\simeq125$, and $0.72$ and $0.57$ at $Re_\tau\simeq550$. Including untyped pairs in a control using $200$ snapshots per Reynolds number leaves the aggregate interface-minus-lobe differences essentially unchanged.

This concentration also depends on the observation frame. It remains pronounced when the observer moves downstream at the prescribed convection velocity, but becomes much weaker at a fixed location (Appendix~\ref{app:robustness}). Tests of temporal alignment also confirm that the peak is associated with the structures at the initial instant of the observed transition (Appendix~\ref{app:temporal_assignment}).

The strength of the interfacial maximum varies with pair arrangement. Streamwise-up pairs generally exhibit the highest surprisal at $Re_\tau\simeq125$, followed by spanwise and wall-normal pairs, whereas streamwise-down pairs exhibit a weaker maximum. This ordering partly reflects the greater near-wall prevalence of streamwise-up configurations, but their advantage also remains in several comparisons within wall-normal bands. A similar ordering occurs at $Re_\tau\simeq550$, particularly above $y^+=20$. The concentration is therefore common to both channels, while its magnitude depends on the organization and wall-normal location of the lobes.

\begin{figure}
    \centering
    \includegraphics[width=0.85\textwidth]{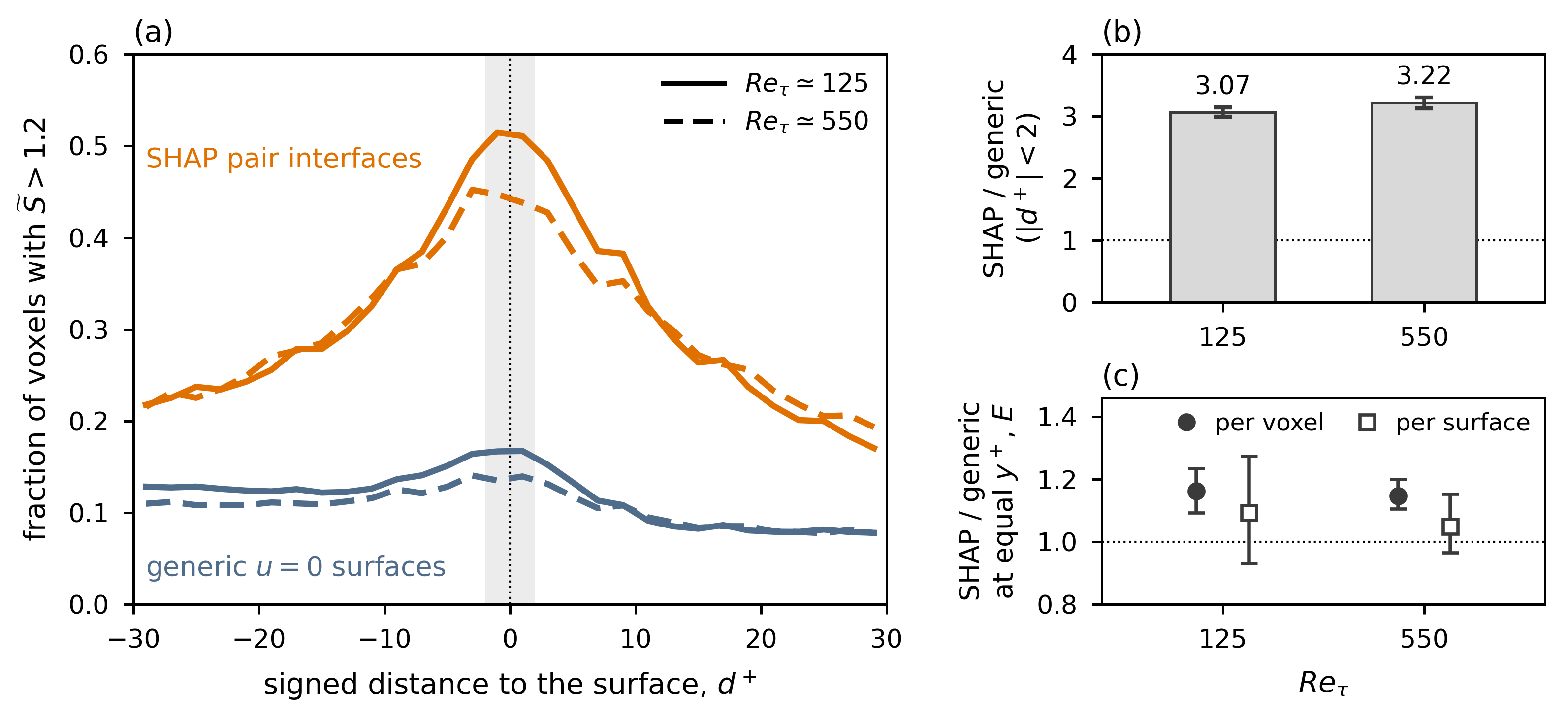}
    \caption{Conditional surprisal near SHAP-selected HS/LS interfaces and generic $u=0$ surfaces at $Re_\tau\simeq125$ and $550$. Generic surfaces are seeded outside the SHAP mask and approximately matched in wall-normal position and area. (a) Fraction of voxels with $\widetilde{S}>1.2$ against signed distance $d^+$ from the surface. Orange and blue curves represent SHAP-selected and generic interfaces, respectively; solid and dashed lines distinguish the Reynolds numbers. The shaded region indicates $|d^+|<2$. (b) Ratio of near-interface high-surprisal fractions, with $95\,\%$ temporal block-bootstrap confidence intervals. (c) Corresponding ratios after matching wall-normal position and fluctuation energy, using voxel weighting (filled circles) or equal weight per surface (open squares). Dotted horizontal lines indicate unity.}
    \label{fig:control_u0}
\end{figure}

The interfacial enhancement raises a further question: are the boundaries within SHAP-selected pairs more surprising than other surfaces separating positive and negative streamwise fluctuations in the same channel? We compare them with generic $u=0$ surfaces constructed from sign changes seeded outside the SHAP mask and matched approximately in surface area and wall-normal location. The comparison reveals a substantial difference [figure~\ref{fig:control_u0}]. At $Re_\tau\simeq125$, the fraction of voxels with $\widetilde{S}>1.2$ within two wall units of the surface is $0.513$ at SHAP pair interfaces, compared with $0.167$ at generic surfaces, a ratio of $3.07$. At $Re_\tau\simeq550$, the corresponding ratio is $3.22$. Thus, strongly surprising velocity transitions are approximately three times as frequent near the interfaces of SHAP-selected pairs as near the generic reference surfaces, a distinction reproduced in both channels.

The contrast is substantially reduced when the surfaces are compared at similar fluctuation energy, indicating that the energetic environment accounts for much of the difference. A smaller excess remains close to the wall in voxel-wise comparisons, while the aggregate contrast is compatible with unity when surfaces are weighted equally. Generic $u=0$ surfaces also show an increase of surprisal towards the sign-change surface itself. The associated conditioning and sensitivity tests are reported in Appendix~\ref{app:increment}.

Finally, the interfacial concentration and the SHAP-only voxel signature of \S\ref{sec:conditional_selection} concern largely different parts of the selected fluid. In $5<y^+<12$, approximately $2.1\,\%$ of the SHAP-only volume lies within eight wall units of an identified pair interface at both Reynolds numbers. Excluding these nearby voxels changes the SHAP-only selection factor by less than $2\,\%$, indicating that the association with surprisal is not confined to the immediate vicinity of these interfaces. The two analyses therefore reveal distinct but complementary properties of the flow emphasized by the forecasting networks.

\subsection{Physical composition of the HS/LS interfaces}
\label{sec:interface_composition}

The internal $u=0$ interface separates the principal HS and LS lobes, but it need not be a region of weak turbulent activity. Although the streamwise fluctuation vanishes there, the fluctuation energy remains $E=v^2+w^2$, and velocity gradients and vortical motions need not vanish. Previous studies of wall turbulence have identified vortex clusters near the boundaries between high- and low-speed streaks, often associated with neighbouring sweep and ejection structures \citep{Dong2017}. These observations suggest a connection between HS/LS interfaces and familiar coherent motions, but they do not establish that the entire interface is vortical. We therefore examine which classical structures intersect the SHAP-selected interfaces and whether their different portions exhibit distinct surprisal signatures.

We partition the reconstructed interface into four mutually exclusive categories: vortical regions within the SHAP mask ($\mathcal H\cap\mathcal V$), other classical regions within the mask ($\mathcal H\cap(\mathcal Q\cup\mathcal L)\setminus\mathcal V$), SHAP-only regions ($\mathcal H\setminus\mathcal C$), and regions outside the SHAP mask. At $u=0$, Q events cannot satisfy their intensity criterion, whereas streaks may intersect the interface through sufficiently strong spanwise fluctuations. The vortical criterion does not impose a fixed sign of $u$. This partition therefore distinguishes the vortically identified portions of the interface from selected fluid that does not satisfy any of the classical criteria.

\begin{figure}
    \centering
    \includegraphics[width=0.95\textwidth]{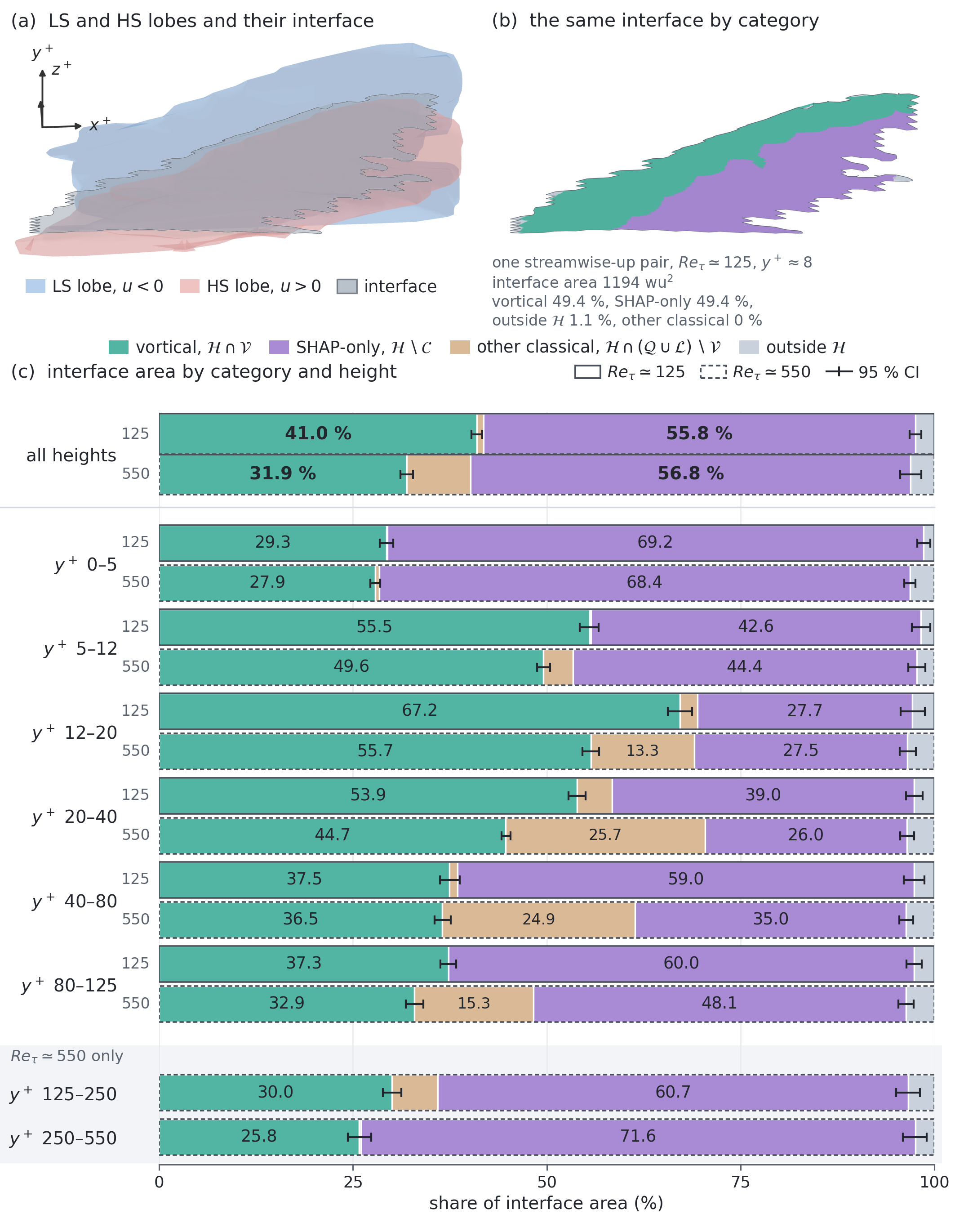}
    \caption{Physical composition of the reconstructed $u=0$ interfaces between the principal HS and LS lobes. (a,b) Illustrative streamwise-up pair at $Re_\tau\simeq125$, showing the principal lobes, their internal interface and its classification into vortical, other classical, SHAP-only and non-selected portions. (c) Fractions of interfacial area assigned to these mutually exclusive categories across the channel and by wall-normal band at $Re_\tau\simeq125$ and $550$. Solid and dashed bar outlines distinguish the Reynolds numbers. Error bars indicate $95\,\%$ block-bootstrap confidence intervals for the vortical and SHAP-only area fractions. Bands above $y^+=125$ occur only at $Re_\tau\simeq550$.}
    \label{fig:interface_composition}
\end{figure}

Figure~\ref{fig:interface_composition}(a,b) illustrates this heterogeneous composition in an individual HS/LS pair. Across the ensemble at $Re_\tau\simeq125$, $41.0\,\%$ of the reconstructed interfacial area lies within vortical structures, while $55.8\,\%$ belongs to SHAP-only [figure~\ref{fig:interface_composition}(c)]. At $Re_\tau\simeq550$, the corresponding fractions are $31.9\,\%$ and $56.8\,\%$. Thus, more than half of the reconstructed interfacial area lies outside the classical coherent-structure masks while remaining within the SHAP mask, despite the substantial contribution of vortical fluid. The interface is therefore neither exclusively vortical nor a boundary devoid of recognizable turbulent activity.

The interfacial composition also changes with wall distance. The vortical fraction is highest between approximately $y^+=5$ and $40$, whereas SHAP-only occupies most of the interfacial area immediately next to the wall and farther from it [figure~\ref{fig:interface_composition}(c)]. At $Re_\tau\simeq550$, non-vortical classical regions, primarily associated with the streak criterion, also contribute appreciably at intermediate heights. The interface therefore crosses different physical environments rather than corresponding to a single conventional structure throughout the channel. This coexistence is consistent with the near-wall self-sustaining cycle, in which quasi-streamwise vortices contribute to streak formation through the lift-up mechanism \citep{HamiltonKimWaleffe1995,JimenezPinelli1999}.

\begin{figure}
    \centering
    \includegraphics[width=0.87\textwidth]{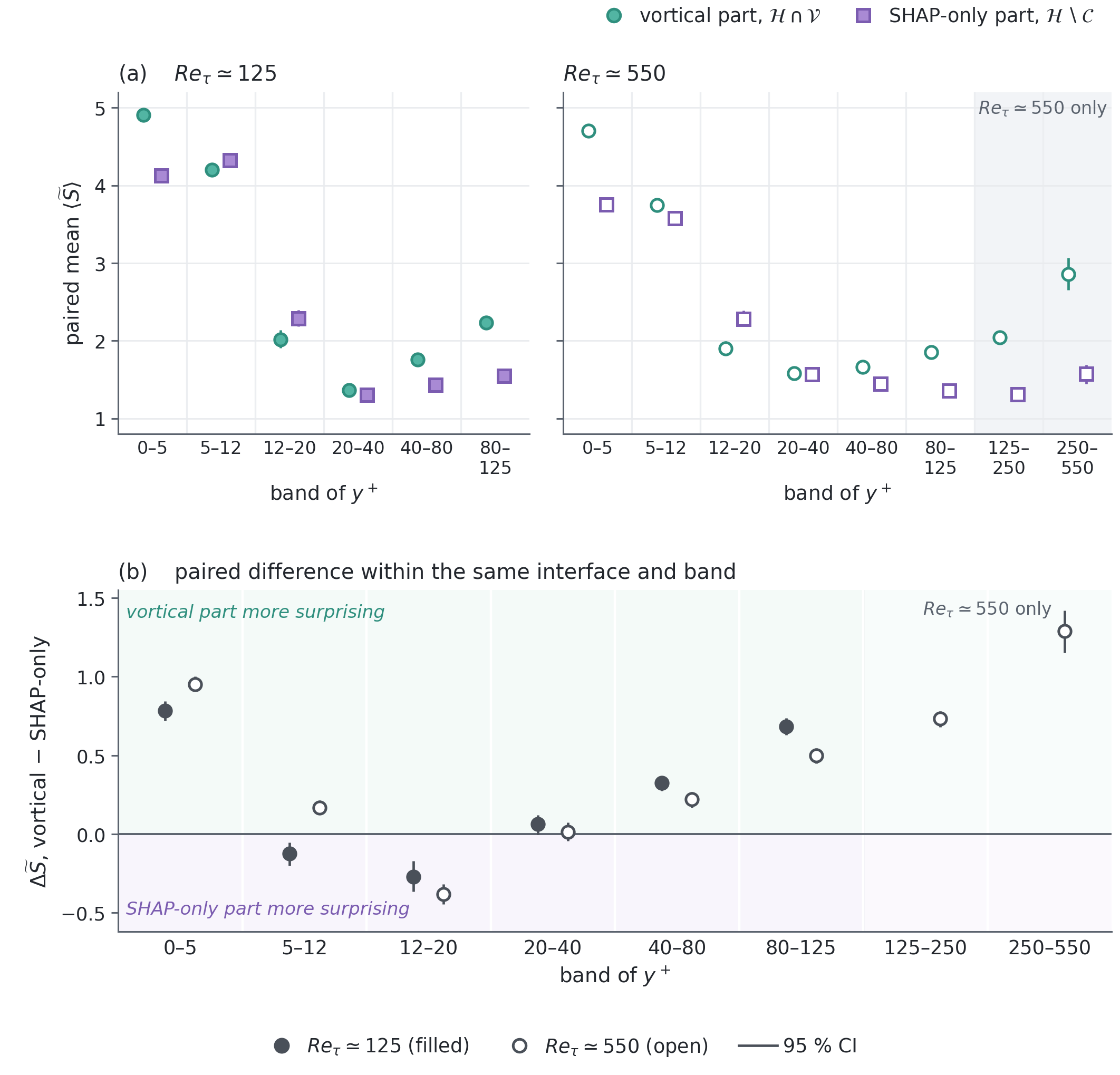}
    \caption{Standardized conditional surprisal in vortical and SHAP-only portions of reconstructed HS/LS interfaces at $Re_\tau\simeq125$ and $550$. (a) Mean $\widetilde{S}$ within wall-normal bands, obtained from area-weighted means over the two portions of each interface and combined using common pair weights. Only interfaces with at least $5$ squared wall units in each category and band are included. (b) Paired difference $\Delta\widetilde{S}_{\mathcal V-\mathrm{SO}}$ between vortical and SHAP-only portions of the same interfaces. Positive values indicate higher surprisal in vortical fluid; negative values favour SHAP-only. Filled and open symbols distinguish $Re_\tau\simeq125$ and $550$, respectively. Error bars indicate $95\,\%$ temporal block-bootstrap confidence intervals.}
    \label{fig:interface_surprisal}
\end{figure}

An important question is whether the surprisal enhancement identified in \S\ref{sec:pair_interfaces} is confined to the vortical portions of these interfaces. We compare the area-weighted mean standardized surprisal of the vortical and SHAP-only portions within the same pair interface and wall-normal band. Figure~\ref{fig:interface_surprisal}(a) shows that both categories exhibit elevated surprisal, particularly near the wall. Their relative magnitudes, however, vary considerably with wall distance, as quantified by the paired difference $\Delta \widetilde{S}_{\mathcal V-\mathrm{SO}}$ [figure~\ref{fig:interface_surprisal}(b)].

Very near the wall ($y^+<5$), the vortical portion has higher mean surprisal than the SHAP-only portion, with paired differences of approximately $0.78$ at $Re_\tau\simeq125$ and $0.95$ at $Re_\tau\simeq550$. A vortical excess also occurs beyond $y^+\simeq40$. In contrast, the two categories exhibit similar surprisal in $20<y^+<40$, despite the substantial vortical area fraction in this region.

In $12<y^+<20$, the ordering reverses: the SHAP-only portions exhibit higher mean standardized surprisal than the vortical portions of the same interface, with paired differences of $-0.27$ and $-0.38$ at the two Reynolds numbers, respectively. A control using $200$ snapshots per Reynolds number shows that much of this contrast is associated with differences in the populations' energy and convected-increment distributions. Restricting the comparison to bins shared by both portions within the same pair reduces the unadjusted differences to $-0.08$ and $-0.14$, respectively. Further matching within these bins gives differences of $-0.04$ and $-0.03$. Thus, the larger contrast in the original populations is concentrated in portions without comparable counterparts in the other category, while a smaller difference persists on the common support.

These observations suggest a connection with the broader organization of wall turbulence, in which vortical motions coexist with shear layers and the deformation or displacement of high- and low-speed streaks. The SHAP-only portions may include dynamically active regions surrounding classical structures or parts of the HS/LS transition that do not exceed their identification thresholds. Their elevated surprisal is consistent with substantial observed velocity changes, although neither the present classification nor surprisal alone establishes the mechanisms producing those changes.

The interfacial results therefore extend the characterization of SHAP-selected fluid beyond conventional coherent-structure boundaries. SHAP-selected HS/LS interfaces contain substantial vortical activity, but their statistically unusual velocity transitions are not restricted to vortically identified regions. In particular, the enhanced surprisal of their non-classical portions provides a complementary statistical signature of selected fluid that remains outside traditional intensity-based criteria, without implying that the forecasting network explicitly evaluates surprisal.

\section{Discussion}
\label{sec:discussion}

\subsection{What distinguishes the fluid selected by SHAP?}
\label{sec:disc_components}

The present results reveal two complementary signatures of the fluid emphasized by predictive attribution in turbulent channel flow. At the voxel level, SHAP selection is dominated by energetic fluid already identified by classical coherent-structure criteria, yet a smaller selected population outside those criteria exhibits unusually high conditional surprisal. At the level of complete structures, SHAP-selected high-/low-speed pairs contain internal interfaces where surprisal concentrates, despite the vanishing streamwise velocity fluctuation. Both signatures occur at $Re_\tau\simeq125$ and $550$. They concern largely different parts of the selected fluid and therefore provide complementary interpretations of its observed evolution.

The first finding helps characterize the incomplete overlap between SHAP structures and classical coherent-structure masks reported by \citet{Cremades2024,Cremades2025}. The energetic SHAP-classical overlap dominates the selected volume, but high fluctuation intensity does not necessarily entail an unusual subsequent velocity transition. SHAP-only regions present a contrasting signature: their association with conditional surprisal persists at comparable energy and wall distance and remains robust to additional controls for proximity to classical structures and local velocity gradients. This suggests that conventional intensity thresholds overlook aspects of the observed evolution of some model-relevant fluid, without identifying the features used internally by the forecasting network.

This result is especially relevant because SHAP-only voxels need not form independent coherent structures. Many occur near classical structures and belong to connected SHAP components that also contain classically identified fluid. They are therefore better understood as portions of selected structures that conventional thresholds do not capture, rather than necessarily constituting a separate family of turbulent motions. Their velocity signatures suggest possible connections with the displacement, deformation and reorganization of nearby structures. For example, the lateral movement or meandering of high- and low-speed streaks may produce appreciable convected velocity changes without requiring particularly large instantaneous streamwise fluctuations. Such mechanisms are consistent with familiar near-wall dynamics, although the present analysis does not identify individual deformation or breakdown events.

The distinction between instantaneous intensity and relevance to subsequent evolution also appears in studies based on perturbations and information theory, although the quantities examined differ from ours. \citet{EncinarJimenez2023} showed that kinetic energy is effective at identifying regions of large absolute perturbation growth in isotropic turbulence, whereas relative amplification depends on additional properties of the flow. In turbulent channels, \citet{OsawaJimenez2024} likewise distinguished the influence of initial perturbation amplitude from the organization of the surrounding flow. Using a different information-theoretic formulation, \citet{ArranzLozanoDuran2024} separated velocity components according to the information they carry about future wall shear stress and found that a substantial fraction of near-wall kinetic energy belongs to the non-informative component. Although these studies address different observables, they illustrate a common limitation of intensity-based descriptions: the magnitude of a velocity fluctuation does not uniquely determine its relationship with subsequent flow evolution.

Conditional surprisal adds a state-conditioned description of observed velocity changes, rather than another measure of instantaneous intensity. Its association with SHAP-only selection is closely tied to convected velocity increments, although surprisal retains some discrimination beyond increment magnitude alone. The stronger association obtained with the convectively moving observer is consistent with a contribution from changes that persist after approximate downstream advection is accounted for, rather than solely from structures passing a fixed location. Because this observer does not track individual material trajectories or remove fluctuating and transverse advection, the result does not by itself identify deformation, reorganization or any particular dynamical mechanism.

SHAP and surprisal nevertheless describe different quantities. SHAP attributes the prediction-error objective of a trained forecasting network, whereas conditional surprisal is evaluated independently from the observed DNS velocity transitions. Their association does not establish that the network explicitly evaluates surprisal or applies successive selection criteria based on energy and statistical unexpectedness. What the comparison does establish is a complementary characterization of selected fluid: regions that are difficult to distinguish through conventional intensity thresholds can exhibit a clear signature in their observed velocity evolution.

\subsection{High-/low-speed interfaces and the organization of wall turbulence}
\label{sec:disc_interfaces}

The second contribution concerns the internal organization of complete SHAP-selected structures. In an adverse-pressure-gradient wing boundary layer, \citet{MolinaCasino2026} identified structures containing adjacent high- and low-speed lobes, with ensemble-averaged attribution concentrated near their separating surface. We find analogous pairs in turbulent channels and, using a descriptor computed from DNS velocity transitions rather than the attribution field, observe enhanced surprisal near their internal $u=0$ interfaces. This provides a complementary statistical perspective on a location previously highlighted by predictive attribution, without requiring the two fields to have identical spatial maxima.

The location of this maximum is physically noteworthy. Although the streamwise fluctuation vanishes on the interface, neither the transverse velocity fluctuations nor the velocity gradients need vanish. Under the zero-fluctuation reference used for expected gradients, the streamwise attribution component is zero at $u=0$, whereas transverse attribution components may remain non-zero. Consequently, the interface can belong to a SHAP-selected structure through its transverse velocity content, while its observed streamwise evolution exhibits statistically unusual convected transitions.

The organization of these interfaces is consistent with familiar elements of wall turbulence. High- and low-speed streaks, ejections and sweeps form spatially organized configurations accompanied by vortical motions and concentrated shear \citep{LozanoDuranFloresJimenez2012,Jimenez2018}. The SHAP-selected pairs exhibit related geometrical features, and substantial portions of their internal interfaces intersect vortical structures. Their classification is nevertheless based on connected SHAP-selected fluid and the relative positions of the principal lobes, rather than on the identification of Reynolds-stress events. The resemblance therefore provides physical context, not an equivalence between the detected objects.

An important extension of this picture is that the interfacial surprisal enhancement is not restricted to vortically identified fluid. At both Reynolds numbers, more than half of the reconstructed interface area belongs to SHAP-only, despite a substantial vortical contribution. The surprisal of these different portions also depends on wall distance. Very near the wall, vortical portions generally exhibit greater mean surprisal, whereas in $12<y^+<20$ the ordering reverses: within the same reconstructed interface and wall-normal band, SHAP-only portions have higher mean standardized surprisal at both Reynolds numbers. This contrast becomes substantially smaller when portions with similar fluctuation energy and convected-increment magnitude are compared, indicating that differences in these quantities contribute to the observed ordering.

This finding suggests that the flow evolution associated with an HS/LS boundary extends beyond the intense regions captured by conventional coherent-structure criteria. The non-classical portions may include surrounding regions of shear, deformation or streak displacement that remain below the thresholds of the classical masks. Such an interpretation is compatible with the known organization of wall turbulence, but it remains a physical hypothesis: neither the vortical classification nor the surprisal measurement identifies which dynamical terms generate the observed velocity changes. In particular, SHAP-only denotes fluid outside the adopted masks, not fluid devoid of vorticity, gradients or coherent motion.

The relative arrangement of the lobes provides another connection with established dynamics. \citet{OsawaJimenez2024} identified dynamically significant near-wall configurations involving high-speed fluid overtaking low-speed fluid. Our streamwise-up pairs share that geometrical ordering and generally exhibit pronounced interfacial surprisal. The correspondence is suggestive, but it does not demonstrate that the lobes approach one another, that the separating surface is compressed or that the structures have the interventional significance established in their perturbation-based study. Their results nevertheless provide a plausible physical setting for the dependence of surprisal on lobe arrangement.

The comparison with generic $u=0$ surfaces provides further context for this result. These surfaces also exhibit enhanced surprisal near velocity-sign changes, indicating that the interfacial peak is not exclusive to SHAP-selected structures. Moreover, much of the raw difference between selected and generic interfaces is associated with their surrounding fluctuation energy: after energy matching, the equal-weight-per-surface comparison is statistically compatible with unity on the retained common support. The interfacial maximum therefore remains a clear feature of the selected HS/LS pairs, while its magnitude relative to generic surfaces is largely related to their different energetic environments.

These findings suggest that conditional surprisal can provide a statistical signature of flow evolution in regions whose organization resembles previously identified turbulent motions, including portions highlighted by SHAP but missed by classical structure criteria. Identifying the physical processes responsible for these transitions, and establishing their dynamical significance, requires further investigation.

\subsection{Implications and limitations}
\label{sec:disc_limitations}

Taken together, the voxel-wise and interface analyses show that no single instantaneous descriptor captures all the statistical properties of the flow emphasized by predictive attribution. Energy accounts for much of the selected volume and for much of the contrast with generic interfaces; conditional surprisal distinguishes an additional aspect of the observed evolution of selected fluid outside conventional masks. The two analyses concern largely different portions of the selected flow and should not be conflated.

Several limitations qualify these conclusions. The two Reynolds numbers involve different DNS configurations, forecasting networks and attribution thresholds, so changes in the measured effect magnitudes cannot be attributed uniquely to Reynolds number. Surprisal is evaluated for the streamwise velocity alone and depends on the prescribed convective observer, which does not follow the fluctuating transverse displacement or deformation of individual structures. Its close association with the convected velocity increment also means that it should not be interpreted as an independent dynamical mechanism or a direct measure of the intrinsic unpredictability of the complete flow. The classical masks depend on their identification criteria and thresholds, while the comparison with generic interfaces does not establish equivalence between the topology and dynamical histories of the selected and reference surfaces.

A further limitation concerns the interpretation of predictive importance. SHAP describes attribution within a trained forecasting model, not the causal influence of a region on the Navier--Stokes evolution. Likewise, conditional surprisal measures the statistical unexpectedness of an observed transition under a chosen reference distribution, not the causal contribution of that transition to the network's prediction. These findings motivate further investigation of the physical processes associated with the selected regions. Tracking SHAP-only regions and high-/low-speed interfaces through time could help assess possible connections with streak displacement, structural deformation, transverse transport and changes in lobe arrangement. In particular, the elevated surprisal of streamwise-up pairs motivates examining their interface evolution and local velocity gradients. Controlled perturbations of selected and matched unselected regions would be needed to determine whether these statistical signatures are associated with greater dynamical influence on the subsequent flow.

Beyond these physical questions, the results suggest a possible methodological application. More generally, assigning a state-conditioned measure of statistical unexpectedness to individual observed transitions allows local flow evolution to be examined independently of a particular forecasting architecture or attribution method. The strong association between SHAP selection and fluctuation energy, together with the additional discrimination provided by information-theoretic descriptors, motivates the search for simpler approximations to computationally expensive attribution maps. One possibility is to investigate interpretable combinations of energy, conditional surprisal and wall distance that reproduce the spatial organization of SHAP-selected fluid without recalculating full attribution fields. Such approximations would require quantitative validation against the original maps and their associated prediction objectives. Moreover, the surprisal used here depends on the subsequently observed velocity field and therefore cannot directly identify important regions before that field is known. A genuinely prospective indicator would require an initial-state-only formulation and independent assessment of its predictive usefulness.

\section{Conclusions}
\label{sec:conclusions}

We have used conditional surprisal to investigate the observed velocity evolution of turbulent flow regions highlighted by predictive attribution. By examining these transitions directly from DNS data, independently of the forecasting network, we provide a statistical interpretation of SHAP-selected fluid that complements conventional coherent-structure criteria. Three main findings emerge.

First, fluctuation energy is the dominant signature of SHAP selection, but it does not tell the whole story. SHAP-only regions exhibit substantially higher conditional surprisal than the selected fluid overlapping classical structures, despite being less energetic. Moreover, the association between surprisal and SHAP-only selection persists at comparable fluctuation energy and wall distance, and remains robust to additional controls. This identifies a distinctive statistical signature of selected fluid that lies outside the classical masks.

Second, conditional surprisal reveals a pronounced spatial organization around the internal $u=0$ interfaces of SHAP-selected high-/low-speed pairs. At both Reynolds numbers, these boundaries exhibit maxima of surprisal in regions where SHAP attribution is also enhanced, providing a connection between predictive importance and the observed evolution of the flow. Although the streamwise velocity fluctuation vanishes at the interface, transverse fluctuations remain, consistent with the complex organization of streaks and vortical motions in wall turbulence. Interestingly, more than half of the reconstructed interfacial area belongs to SHAP-only, whose mean surprisal exceeds that of the vortical portions in part of the buffer layer. This difference becomes much smaller when comparing portions with similar fluctuation energy and convected increments, suggesting that their different local flow conditions contribute substantially to the observed contrast. A similar distinction emerges when comparing SHAP-selected interfaces with generic $u=0$ surfaces: while the former initially exhibit much higher surprisal, most of this excess is associated with differences in surrounding fluctuation energy, and the contrast is no longer statistically significant when surfaces of comparable energy are given equal weight. Together, these findings show how conditional surprisal complements SHAP by revealing the organization of unusual velocity transitions around coherent flow interfaces, while helping to distinguish their spatial signatures from the influence of the surrounding turbulent fluctuations.

Finally, these findings demonstrate the value of information-theoretic analysis for interpreting the flow highlighted by explainable deep learning. Although the observed associations do not establish causal influence or reveal the forecasting network's internal mechanisms, they motivate further investigation of the physical processes behind unusual velocity transitions, including temporal tracking and controlled perturbations. They also open the possibility of developing simpler, interpretable flow descriptors that approximate SHAP-selected regions without requiring full attribution calculations.


\begin{bmhead}[Funding]
S.H. acknowledges financial support from the Generalitat
Valenciana through the PROMETEO programme
(SHARP project, grant CIPROM/2024/35).
\end{bmhead}

\begin{bmhead}[Declaration of interests]
The authors report no conflict of interest.
\end{bmhead}

\bibliographystyle{jfm}
\bibliography{bibliografia_JFM_unificada}

\appendix
\begin{appen}
\FloatBarrier
\section{Numerical assessment of the conditional-surprisal estimator}
\label{app:estimator_validation}

\subsection{Histogram resolution and Gaussian benchmark}
\label{app:histogram_validation}

We first examine the numerical stability of the conditional-surprisal estimator by varying the histogram resolution and comparing the finite-window approach with an unregularized density histogram and a histogram of discrete probability masses. Here, $N_b$ denotes the number of histogram bins, while $\kappa$ controls the width of the regularization window relative to the velocity fluctuations within each wall-parallel plane. For three representative planes at $Re_\tau\simeq125$, increasing $N_b$ from $120$ to $400$ changes the mean estimated surprisal by only $0.004$--$0.018$ nats with finite-window regularization, whereas the unregularized density estimator exhibits relative changes of $3$--$8\,\%$ [figure~\ref{fig:validation_A1}(a)]. The estimate based on discrete probability masses also varies with bin width, making it unsuitable for direct comparison across histogram resolutions.

To assess accuracy against a known reference, we also consider a Gaussian benchmark. For jointly Gaussian velocity pairs $(X,Y)$ with unit variances and prescribed correlation $\rho$, the conditional differential entropy is known analytically:

\begin{equation}
    h(Y\mid X)=\tfrac12\log\!\left[2\pi e(1-\rho^2)\right].
    \label{eq:gaussian_entropy}
\end{equation}

Since the expected surprisal equals the conditional differential entropy when the underlying conditional density is known, this benchmark allows us to evaluate how closely the estimator reproduces the theoretical value. For correlations $\rho\in\{0,0.3,0.5,0.7,0.9,0.95\}$, using four realizations per correlation, the adopted parameters $\kappa=0.1$ and $N_b=120$ yield mean and worst-case absolute errors of $4.9\times10^{-3}$ and $7.3\times10^{-3}$ nats, respectively [figure~\ref{fig:validation_A1}(b,c)]. Although $\kappa=0.15$ and $N_b=200$ produce slightly smaller errors, the adopted configuration balances estimation accuracy with localization in velocity space.

However, the DNS correlations reach approximately $\rho=0.99$--$0.997$ in the outer region at $Re_\tau\simeq550$, beyond the original Gaussian benchmark. Extending the test to these values reveals greater sensitivity at high correlations: with $N_b=120$, the Spearman correlation between estimated and exact pointwise surprisal decreases from $0.95$ at $\rho=0.98$ to $0.75$ at $\rho=0.997$, but increases to $0.94$ at the latter correlation when $N_b=300$. This limitation is mainly relevant to the outer region at $Re_\tau\simeq550$. The principal SHAP-surprisal associations examined here occur closer to the wall, where their stability with respect to histogram resolution is assessed in Appendix~\ref{app:dns_validation}.

\begin{figure}
    \centering
    \includegraphics[width=\textwidth]{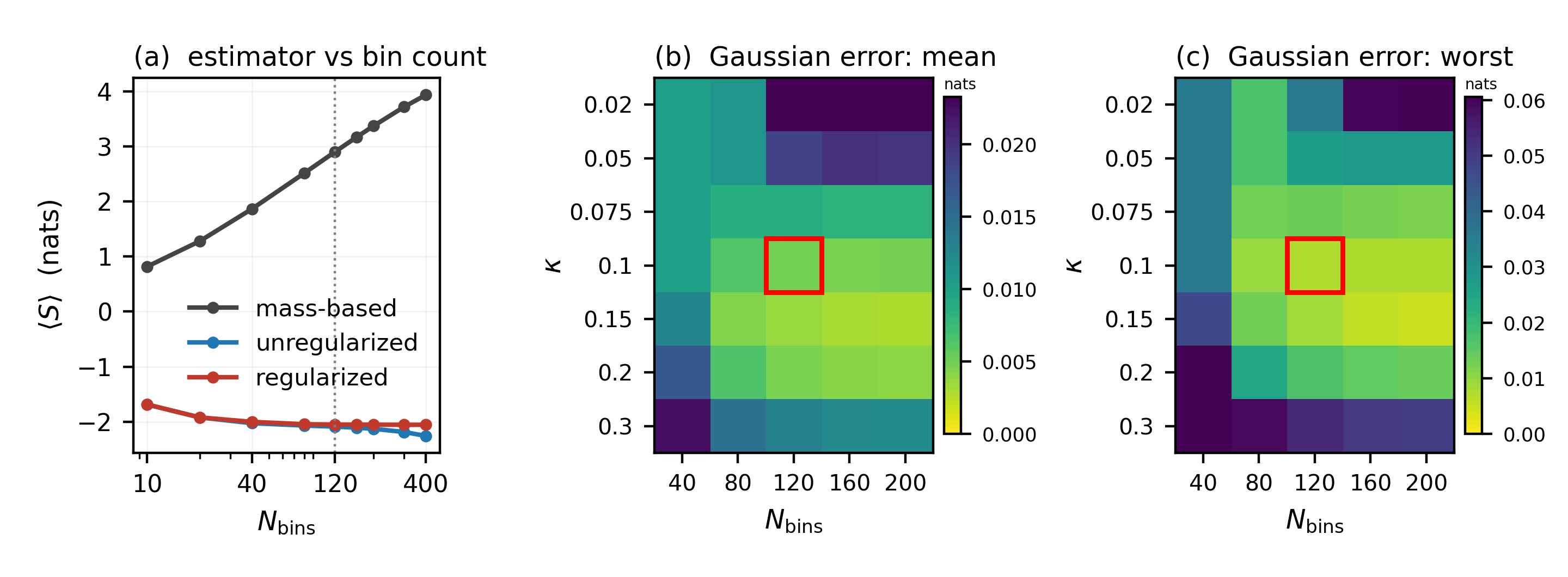}
    \caption{Numerical assessment of the conditional-surprisal estimator. (a) Mean surprisal versus histogram resolution for estimators based on discrete probability masses, unregularized densities and finite-window regularization, illustrated at $y^+\simeq14.5$ for one DNS snapshot pair at $Re_\tau\simeq125$. (b,c) Mean and maximum absolute errors $|\langle S\rangle-h(Y\mid X)|$, in nats, for the Gaussian benchmark as functions of $N_b$ and $\kappa$, averaged over four realizations. The red boxes indicate the adopted parameters $(N_b,\kappa)=(120,0.1)$.}
    \label{fig:validation_A1}
\end{figure}

\subsection{DNS sensitivity and spatial stability}
\label{app:dns_validation}

We next examine how the choice of histogram resolution and regularization affects the surprisal estimates obtained from the DNS at both $Re_\tau\simeq125$ and $550$, using the convective observer adopted throughout the study. As shown in figure~\ref{fig:validation_A2}(a,d), the distribution of raw surprisal varies with the regularization width, as expected when estimating a continuous probability density. However, changes in the estimated values do not necessarily imply changes in where high surprisal occurs. We therefore compare the spatial fields obtained with different histogram resolutions against a reference estimate using $400$ bins.

Besides measuring their spatial correlation, we examine whether the same voxels remain among those with the highest surprisal. For this purpose, we use the Jaccard index

\begin{equation}
    J_N=\frac{|\mathcal T_N\cap\mathcal T_{400}|}{|\mathcal T_N\cup\mathcal T_{400}|},\qquad
    \mathcal T_N=\{\boldsymbol x:S_N(\boldsymbol x)\ge q_{0.99}(S_N)\},
    \label{eq:jaccard}
\end{equation}

where $S_N$ is the surprisal estimated with $N$ bins and $q_{0.99}$ denotes its $99$th percentile. The index measures the overlap between the sets of highest-surprisal voxels, ranging from zero for no overlap to unity when the sets are identical. Figure~\ref{fig:validation_A2}(b,e) shows that the spatial fields remain strongly correlated across the tested resolutions, while the Jaccard overlap is typically $0.86$--$0.90$. It decreases to approximately $0.82$ above $y^+=250$ at $Re_\tau\simeq550$, indicating somewhat greater sensitivity when identifying the most extreme surprisal values in the outer region. This is one reason why the conditional-selection analysis uses the median surprisal of each wall-parallel plane and snapshot, rather than a threshold based on the extreme tail.

More importantly, we examine whether these variations affect the observed relationship between surprisal and SHAP-only selection. We repeat the comparison using the energy-conditioned selection factor $R_{S\mid E}$, which measures how much more likely SHAP-only selection is for above-median than for below-median surprisal at comparable fluctuation energy and wall distance. In figure~\ref{fig:validation_A2}(c,f), each factor is divided by the value obtained with the adopted parameters $(N_b,\kappa)=(120,0.1)$, so values close to unity indicate that the selection factor changes little. In the main near-wall band, $5<y^+<12$, increasing $N_b$ from $120$ to $200$ changes this factor by only $1.9\,\%$ at $Re_\tau\simeq125$ and $0.6\,\%$ at $Re_\tau\simeq550$; the largest changes across the examined bands are $4.4\,\%$ and $2.5\,\%$, respectively. Changing the regularization parameter $\kappa$ between $0.075$ and $0.15$ produces somewhat larger variations in the factor but preserves the near-wall association between high surprisal and SHAP-only selection. A separate test with $50$ snapshot pairs per Reynolds number further shows that increasing $N_b$ from $120$ to $300$ changes the near-wall selection factors by no more than approximately $4\,\%$, without changing the ordering of the population-mean surprisal values. Thus, despite the greater sensitivity of individual surprisal estimates in the outer region at $Re_\tau\simeq550$, the main near-wall results are stable over the tested estimator settings.

Finally, we assess whether truncation of the regularization windows at the histogram boundaries affects the results. Using $50$ snapshot pairs per Reynolds number and $11$--$12$ representative wall-normal planes, we compare the original estimator with a mass-conserving boundary correction, retaining the same convective observer. The correction changes the estimated surprisal for at most $0.5\,\%$ of the observations in any examined plane, with a maximum difference of approximately $0.5$ nats. No more than $1.1\,\%$ of the voxels in the highest surprisal percentile are affected, and the SHAP-only selection factor changes by less than $0.1\,\%$ on the available attribution pairs. Boundary truncation therefore has a negligible influence on the population statistics considered here.

\begin{figure}
    \centering
    \includegraphics[width=0.85\textwidth]{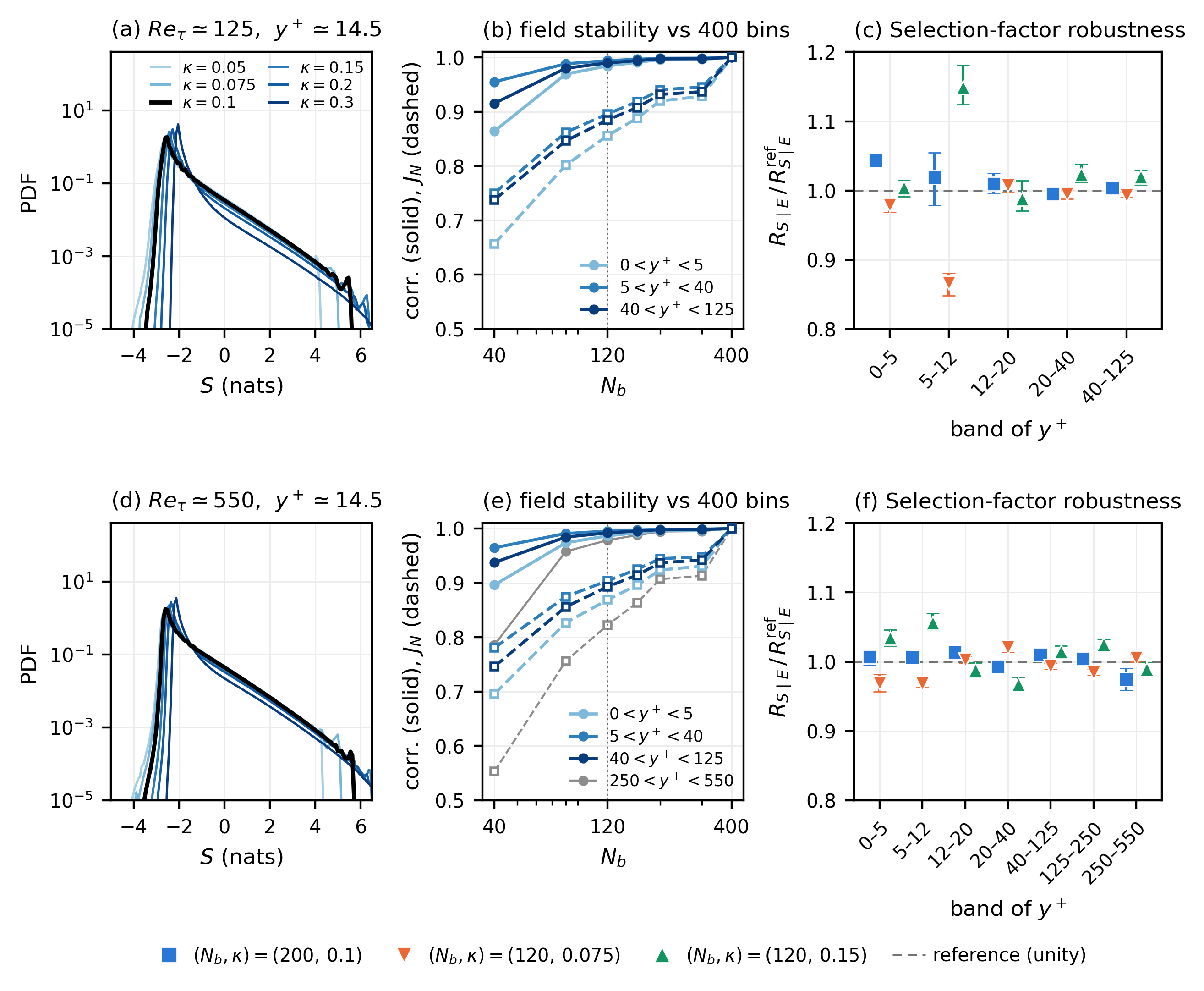}
    \caption{Sensitivity of the conditional-surprisal estimator at $Re_\tau\simeq125$ (upper row) and $550$ (lower row). (a,d) Raw-surprisal distributions for different $\kappa$ at $y^+\simeq14.5$. (b,e) Spatial correlation (solid lines) and Jaccard overlap (dashed lines) relative to a $400$-bin reference. (c,f) Energy-conditioned SHAP-only selection factors normalized by the values obtained with $(N_b,\kappa)=(120,0.1)$; unity indicates no change from the reference result. Panels (b,c,e,f) use $15$ snapshot pairs equispaced over the analysed record. Error bars in (c,f) indicate $95\,\%$ confidence intervals from $1000$ paired bootstrap resamples of these snapshots; panels (b,e) show medians without confidence intervals.}
    \label{fig:validation_A2}
\end{figure}

\FloatBarrier
\section{Contributions, enrichment and comparison with the classical families}
\label{app:conditional_selection}

\subsection{Contributions to plane-wide totals}
\label{app:contributions}

A population may exhibit high local surprisal or fluctuation energy while contributing little to a plane-wide total if its volume fraction is small. We distinguish these quantities by defining the energy contribution of population $A$ as
\begin{equation}
    P_E(A;y)=100\,
    \frac{\displaystyle\sum_t\sum_{p\in\Pi(y)}\sum_{\boldsymbol{x}\in A(t)\cap\mathcal P_p}E(\boldsymbol{x},t)}
         {\displaystyle\sum_t\sum_{p\in\Pi(y)}\sum_{\boldsymbol{x}\in\mathcal P_p}E(\boldsymbol{x},t)},
    \label{eq:energy_contribution}
\end{equation}
and defining $P_\Phi(A;y)$ analogously with $\Phi$ replacing $E$. Because standardized surprisal has a zero reference mean and can take either sign, we instead define its signed contribution per plane voxel as
\begin{equation}
    C_S(A;y)=
    \frac{\displaystyle\sum_t\sum_{p\in\Pi(y)}\sum_{\boldsymbol{x}\in A(t)\cap\mathcal P_p}\widetilde{S}(\boldsymbol{x},t)}
         {\displaystyle\sum_t\sum_{p\in\Pi(y)}|\mathcal P_p|}.
    \label{eq:signed_surprisal_contribution}
\end{equation}
With the voxel occupancy
\begin{equation}
    f_A(y)=\frac{\displaystyle\sum_t\sum_{p\in\Pi(y)}|A(t)\cap\mathcal P_p|}
                 {\displaystyle\sum_t\sum_{p\in\Pi(y)}|\mathcal P_p|},
    \label{eq:population_occupancy}
\end{equation}
this contribution satisfies
\begin{equation}
    C_S(A;y)=f_A(y)\langle \widetilde{S}\rangle_A(y).
    \label{eq:contribution_occupancy}
\end{equation}

Figure~\ref{fig:structure_contributions} illustrates the distinction between local intensity and aggregate contribution. Near $y^+\simeq15$, streaks carry more than $40\,\%$ of the fluctuation energy and approximately $37$--$38\,\%$ of the attribution magnitude, although their mean standardized surprisal is below the plane reference. SHAP-only voxels make a positive contribution to signed surprisal but occupy too little volume to dominate plane-wide totals. The classical families overlap, so their contributions are not additive.

\begin{figure}
    \centering
    \includegraphics[width=0.9\textwidth]{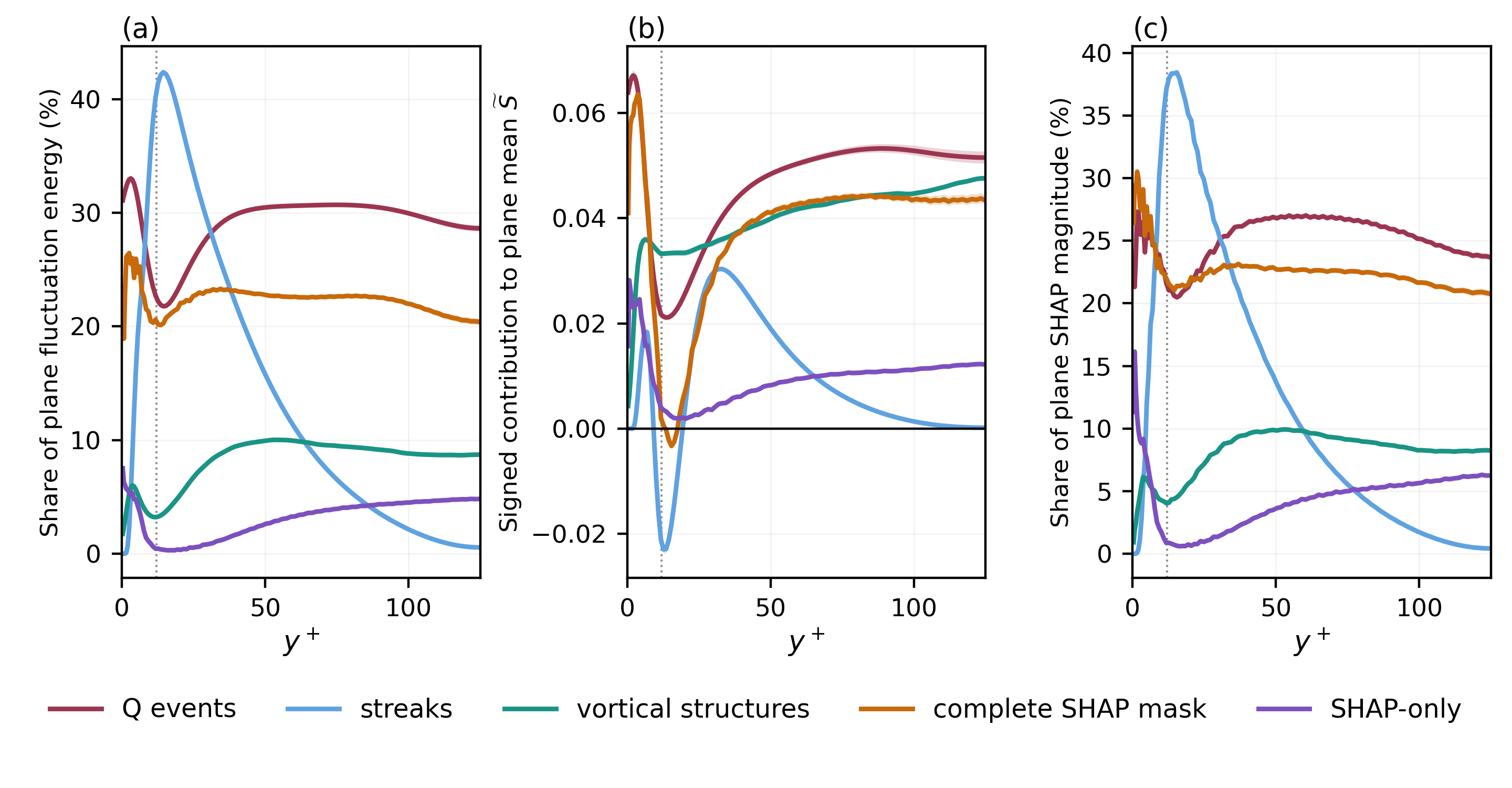}
    \caption{Aggregate contributions of classical structure families and SHAP populations at $Re_\tau\simeq125$. (a) Fraction of total fluctuation energy $P_E$; (b) signed surprisal contribution per plane voxel, $C_S=f_A\langle\widetilde{S}\rangle_A$; and (c) fraction of total plane-wide attribution magnitude $P_\Phi$. Classical structure families can overlap, so their contributions are not additive. Shaded regions indicate $95\,\%$ temporal block-bootstrap confidence intervals; the vertical reference marks $y^+=12$.}
    \label{fig:structure_contributions}
\end{figure}

\subsection{Surprisal enrichment and comparison with classical families}
\label{app:enrichment}
\label{app:families}

The preceding analysis describes the contribution of each population to the plane-wide totals. A complementary measure is the prevalence of high-surprisal voxels within each population at comparable fluctuation energy and wall distance. Unlike the selection-probability analysis in \S\ref{sec:conditional_selection}, which measures how SHAP selection varies with surprisal, this comparison measures the enrichment of above-median surprisal within population $A$:

\begin{equation}
    R_A(k,m)=\frac{P(S_>\mid A,k,m)}{P(S_>\mid k,m)},
    \label{eq:enrichment}
\end{equation}
where $S_>$ denotes surprisal above the plane-wise median and $k$ and $m$ denote wall-normal and energy strata. Thus, $R_A>1$ indicates that high-surprisal voxels are more prevalent within population $A$ than in the corresponding reference stratum; it is not a probability ratio for SHAP selection. The upper bound within a given stratum is $1/P(S_>\mid k,m)$, and therefore depends on the reference fraction in that stratum.

At $Re_\tau\simeq125$ and $5<y^+<12$, $91.4\,\%$ of SHAP-only volume has above-median surprisal. Within the energy strata retained under the sampling criterion of \S\ref{sec:conditional_selection}, the corresponding fraction is approximately $0.89$--$0.97$, compared with $0.45$--$0.70$
among all voxels in the same strata, giving enrichment factors of approximately $1.39$--$2.02$. These values approach the attainable enrichment in strata where the reference high-surprisal fraction is close to one half. Among the classical masks, vortical structures exhibit the enrichment most similar to SHAP-only near the wall, although SHAP-only exceeds the vortical population across the supported energy strata. Q events are less enriched, and streaks have enrichment near or below unity. The distinction between SHAP-only and vortical fluid diminishes farther from the wall. These comparisons complement, but do not replace, the selection-probability analysis in \S\ref{sec:conditional_selection}.

\FloatBarrier
\section{Robustness to the convective observer}
\label{app:robustness}

Conditional surprisal depends on the frame in which a velocity transition is observed. We assess this dependence by comparing the prescribed convective observer $S_3$, with displacement $\ell(y)=U_{\mathrm{conv}}(y)\Delta t$, with an observer $S_2$ moving at the local mean velocity, $\ell(y)=\overline U(y)\Delta t$, and a fixed observer $S_1$, for which $\ell=0$. Each surprisal field is standardized using its own wall-normal reference distribution. 

We first examine how the choice of frame affects the velocity change attributed to downstream advection. For a trial streamwise frame velocity $c$, which need not equal the local mean velocity, we compare the mean-square increment with its fixed-frame counterpart:
\begin{equation}
    F(c)=\frac{\langle[U(x+c\Delta t,y,z,t_{s+1})-U(x,y,z,t_s)]^2\rangle}
               {\langle[U(x,y,z,t_{s+1})-U(x,y,z,t_s)]^2\rangle},
    \label{eq:frame_increment}
\end{equation}
Using representative snapshot pairs, the prescribed convection speed gives $F(c)\simeq0.13$--$0.15$ near the wall, compared with approximately $0.66$ for the local mean-velocity observer [figure~\ref{fig:observer_frame}(a)]. It also lies within $0.03$ of the minimum obtained by varying $c$ over the tested range. This supports the prescribed frame as a practical approximation for reducing the contribution of advection, without identifying a unique propagation speed for every structure. The rank correlations in figure~\ref{fig:observer_frame}(b) show how the relation between surprisal and velocity increments changes across these observational frames.

\begin{figure}
    \centering
    \includegraphics[width=0.85\textwidth]{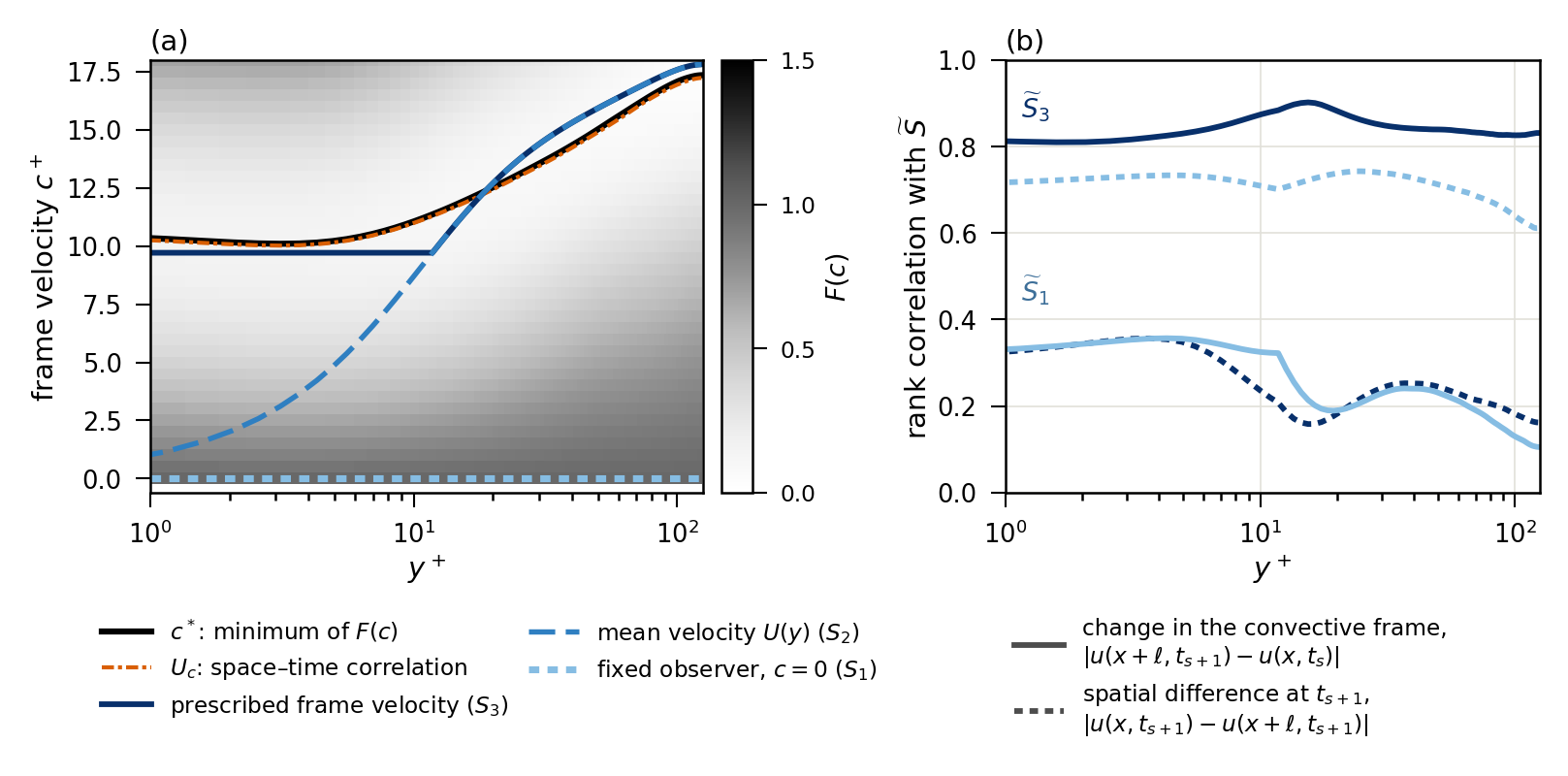}
    \caption{Dependence of velocity-change statistics on the observation frame at $Re_\tau\simeq125$. (a) Mean-square velocity increment relative to the fixed observer, $F(c)$, as a function of wall distance and frame velocity $c^+=c/u_\tau$. The prescribed frame velocity ($S_3$), local mean velocity ($S_2$) and fixed observer ($S_1$) are compared with the speed $c^*$ minimizing $F(c)$ and an independently estimated convection velocity $U_c$ obtained from two-time streamwise correlations. (b) Within-plane rank correlations of standardized surprisal with the convected temporal increment $|u(x+\ell,t_{s+1})-u(x,t_s)|$ (solid) and the spatial difference $|u(x,t_{s+1})-u(x+\ell,t_{s+1})|$ at the final instant (dotted). The analysis uses $40$ representative snapshot pairs.}
    \label{fig:observer_frame}
\end{figure}

The association between surprisal and SHAP-only selection is then evaluated using the alternative observers. In $5<y^+<12$, the energy-conditioned selection factors are $10.9$ for $S_3$, $12.1$ for $S_2$ and $4.6$ for $S_1$, using the same $1000$ snapshot pairs as the proximity analysis in Appendix~\ref{app:margins}. Both moving frames therefore retain a strong association, while the fixed frame yields a weaker, but still positive, selection factor.

The interfacial concentration is more sensitive to this choice of frame. Repeating the analysis of \S\ref{sec:pair_interfaces} on $1500$ snapshots gives interface-minus-lobe differences of $0.98$ (95\,\% confidence interval (CI): $[0.95, 1.01]$) for the prescribed observer $S_3$ and $0.79$ (95\,\% CI $[0.77, 0.82]$) for the local-mean observer $S_2$. By contrast, the fixed observer $S_1$ gives $-0.02$ (95\,\% CI $[-0.04, 0.00]$) [figure~\ref{fig:observer_interface}]. Thus, the pronounced interfacial maximum is preserved in both moving frames but not in the fixed frame. It characterizes velocity transitions measured relative to a convectively moving observer, rather than an observer-independent property of the interface.

\begin{figure}
    \centering
    \includegraphics[width=0.85\textwidth]{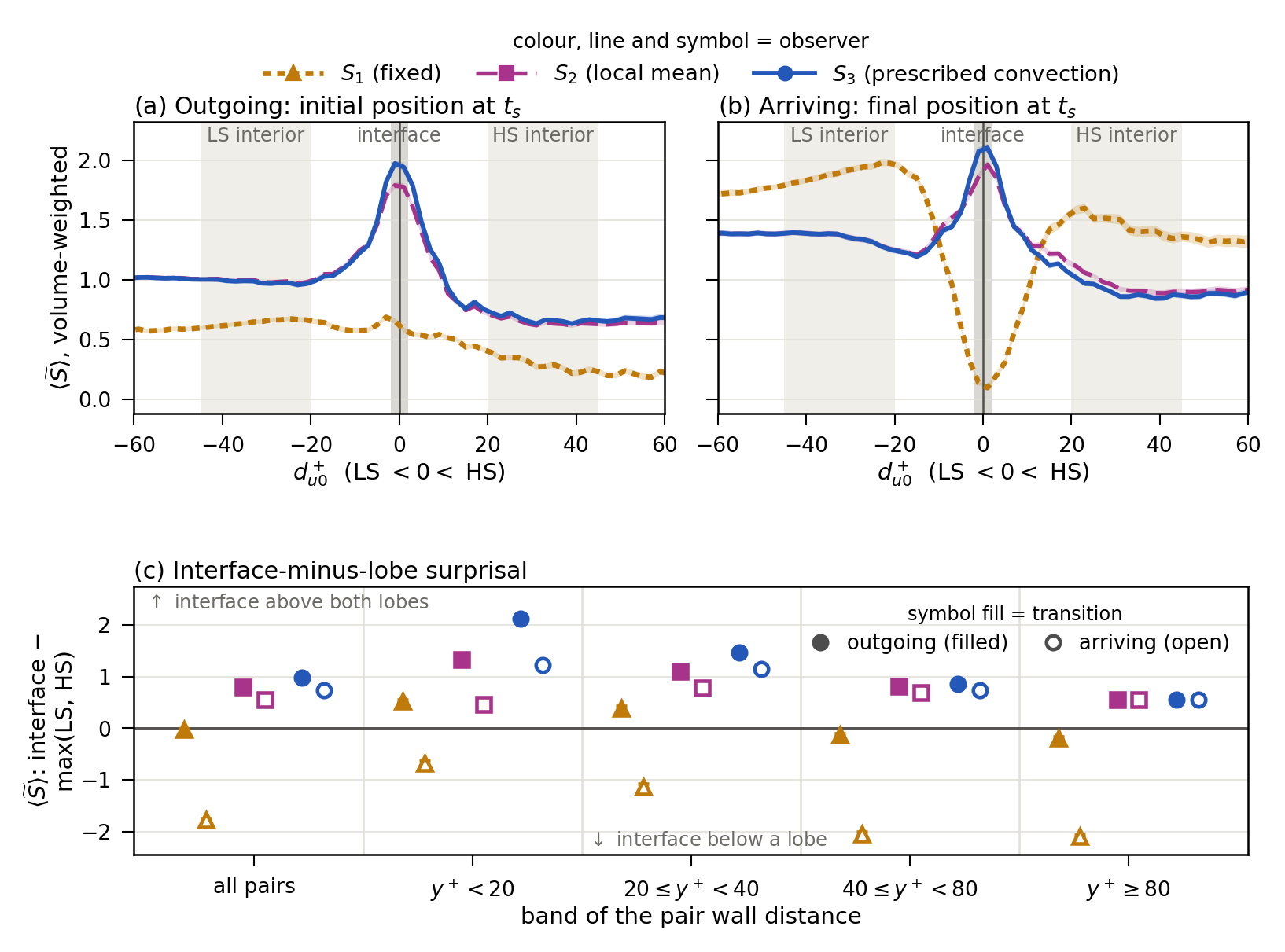}
    \caption{Surprisal concentration near HS/LS interfaces under three observers at $Re_\tau\simeq125$. Each observer's surprisal is standardized using its own wall-normal reference distribution. (a) Volume-weighted mean standardized surprisal $\widetilde S$ against signed interface distance $d_{u0}^+$ for outgoing transitions assigned to their initial position. (b) Corresponding profiles for arriving transitions assigned to their final position. In both panels, grey vertical bands mark the interface ($|d_{u0}^+|\leq2$) and the LS and HS lobe interiors ($20\leq|d_{u0}^+|\leq45$). (c) Interface-minus-lobe difference for the fixed ($S_1$), local-mean ($S_2$) and prescribed convective ($S_3$) observers, across wall-normal bands. Colours distinguish observers, while filled and open symbols indicate outgoing and arriving transitions, respectively. Shading around the profiles and error bars indicate $95\,\%$ temporal block-bootstrap confidence intervals.}
    \label{fig:observer_interface}
\end{figure}

\FloatBarrier
\section{Temporal and geometrical robustness of the pair-interface analysis}
\label{app:temporal}

\subsection{Temporal assignment of surprisal}
\label{app:temporal_assignment}

Conditional surprisal describes a transition between two consecutive snapshots, whereas each SHAP structure is defined at a single instant $t_s$. Their spatial comparison therefore depends on assigning the observed transition to a position corresponding to the structure geometry. In the main analysis, the outgoing transition from $t_s$ to $t_{s+1}$ is assigned to its initial position at $t_s$. We also consider the arriving transition from $t_{s-1}$ to $t_s$, assigned to its final position at the same instant. Both assignments register the transition with the same SHAP pair geometry. The comparison uses $1500$ snapshots at $Re_\tau\simeq125$ and does not require re-estimating the conditional densities.

The interfacial enhancement is recovered with either assignment [figure~\ref{fig:temporal}]. The interface-minus-lobe difference is $0.98$ (95\,\% CI $[0.95, 1.01]$) for outgoing transitions and $0.74$ (95\,\% CI $[0.70, 0.77]$) for arriving transitions. The smaller difference for arriving transitions reflects higher surprisal within the lobes, rather than a disappearance of the interfacial maximum. With equal weight per pair, the respective differences are $1.09$ and $0.99$, again preserving the enhancement near the interface.

As a separate spatial-registration control, we assign the transitions to the midpoints of their convective displacements instead of their initial or final positions. These midpoints do not coincide with the observation positions at the instant of the SHAP geometry. The resulting profiles show a weaker interfacial concentration, particularly away from the wall, consistent with a displacement of the surprisal pattern relative to the reconstructed boundary. The comparison thus supports robustness across the two temporally consistent assignments, while showing that spatial registration matters.

\begin{figure}
    \centering
    \includegraphics[width=0.85\textwidth]{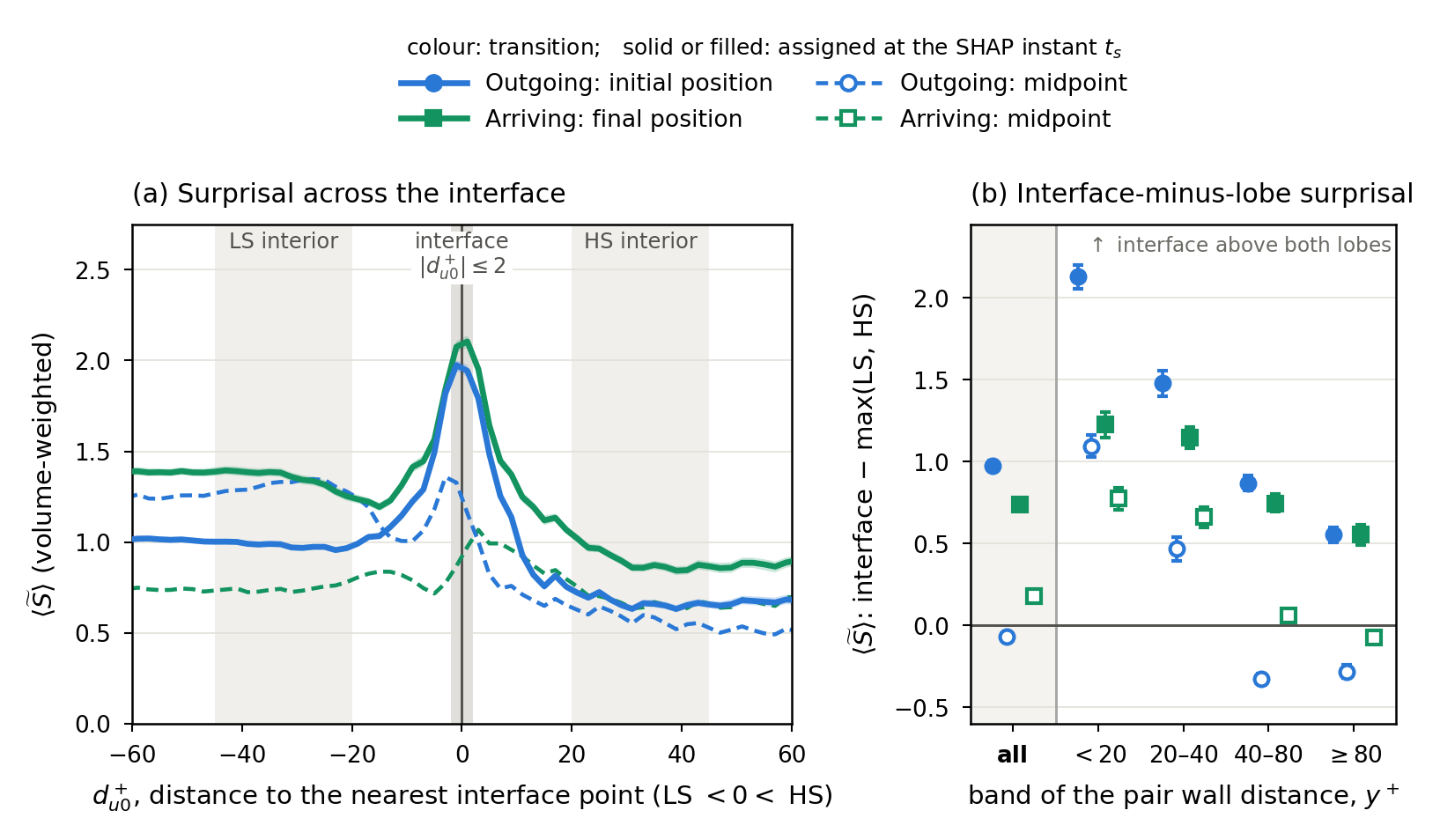}
    \caption{Sensitivity of surprisal concentration near HS/LS interfaces to temporal and spatial assignment at $Re_\tau\simeq125$. Outgoing and arriving transitions are assigned to their initial and final positions, respectively, aligning both with the SHAP structures at $t_s$. As a spatial-registration control, surprisal is alternatively assigned to the midpoint of the convective displacement. (a) Volume-weighted mean standardized surprisal $\widetilde S$ as a function of signed distance $d_{u0}^+$ from the interface. Grey bands mark the interfacial region ($|d_{u0}^+|\leq2$) and the LS and HS lobe interiors ($20\leq|d_{u0}^+|\leq45$). (b) Interface-minus-lobe surprisal differences for all pairs and by wall-normal band. Colours distinguish outgoing and arriving transitions, while filled and open symbols indicate assignments at the SHAP instant and at the spatial midpoint, respectively. Shading around the profiles and error bars indicate $95\,\%$ temporal block-bootstrap confidence intervals.}
    \label{fig:temporal}
\end{figure}

\subsection{Sensitivity to the interface-distance construction}
\label{app:interface_distance}

We also assess whether the interfacial maximum depends on the distance construction. The signed distance $d_{u0}^+$ is measured from each voxel to the nearest interpolated $u=0$ crossing between the principal HS and LS lobes of the same SHAP structure. Using $60$ randomly selected snapshots, we compare distances to interpolated crossing points and to a triangulated interface, while examining periodic boundaries and smaller secondary lobes. These alternatives changed the pooled interface-minus-lobe difference by at most $0.02$, and by at most $0.06$ within individual wall-normal bands. They did not displace the maximum of the pooled profile under either temporally consistent assignment. Excluding secondary-lobe voxels changes the reference population rather than the interface geometry, slightly increasing the contrast. These results indicate that the interfacial maximum is not determined by the particular distance construction used in the main analysis.

\subsection{Geometric registration of the interfacial maximum}
\label{app:geometric_registration}

The previous tests concern the temporal assignment and reconstruction of the internal interface. We also assess whether the surprisal maximum is preferentially registered to that boundary, rather than to the overall location of its SHAP structure. For each pair, we compare an internally standardized profile against signed distance from the reconstructed interface with a profile against streamwise displacement from the structure centroid. Both use the same $\pm200$-wall-unit window, approximately $8$-wall-unit bins and the same temporal block-bootstrap procedure. Profiles are standardized separately within each pair, as described in \S\ref{sec:pairs}, and then averaged with equal weight per pair, so that the comparison emphasizes spatial localization rather than differences in absolute surprisal between structures.

The interface-referenced profile has a pronounced maximum near zero signed distance, while the centroid-referenced profile shows a much weaker enhancement [figure~\ref{fig:geometric_registration}]. These two profiles use different distance coordinates, so we additionally compare both reference locations using the same streamwise coordinate and pairs at $Re_\tau\simeq125$. In this control, the peak rises from approximately $0.17$ for the structure centroid to $0.31$ for the interface, compared with $0.98$ when signed distance from the interface is used. Independently circularly shifting the pair profiles before averaging also removes the sharp interfacial maximum. The enhancement is therefore preferentially registered to the HS/LS interface, although its apparent concentration is influenced by the distance coordinate.

\begin{figure}
    \centering
    \includegraphics[width=0.85\textwidth]{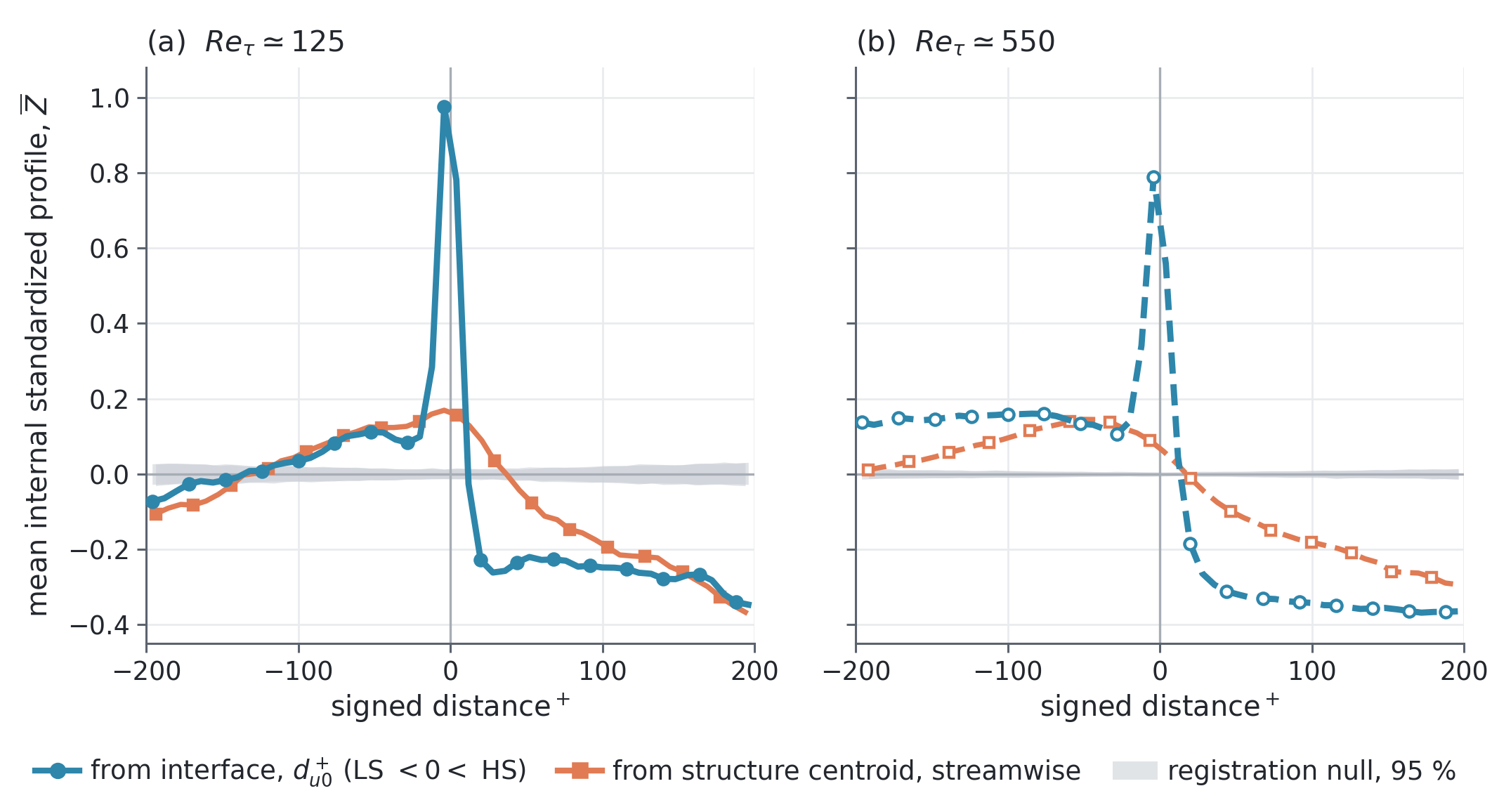}
    \caption{Geometric registration of the surprisal maximum within SHAP-selected HS/LS pairs. Equal-weight-per-pair mean internally standardized surprisal profiles $\overline Z$ referenced to signed distance from the reconstructed interface (blue) and to streamwise distance from the SHAP-structure centroid (coral). Panels (a,b) correspond to $Re_\tau\simeq125$ and $550$, using $1500$ and $3417$ snapshots, respectively. The shaded region indicates the pointwise $95\,\%$ interval of the geometric registration null obtained by independently circularly shifting the resolved profiles before averaging.}
    \label{fig:geometric_registration}
\end{figure}

\FloatBarrier
\section{Surprisal and the convected velocity increment}
\label{app:increment}

\subsection{Relationship between surprisal and velocity increments}

Conditional surprisal depends on both the initial velocity and its observed change. We quantify how much of the spatial variability of standardized surprisal is associated with the magnitude of the convected increment alone, and how that description changes when the initial state and signed increment are included. For each transition, we define $\Delta U_\ell=Y-X$ and its normalized magnitude $D_1=|\Delta U_\ell|/\sigma_\Delta(y,t_s)$, where $\sigma_\Delta$ is the plane-wise root-mean-square increment. Using $1500$ snapshot pairs per Reynolds number, we quantify the explained variance through the volume-weighted correlation ratio $\eta^2$, calculated from conditional bin means within wall-normal bands. The analysis uses $40$ bins for $D_1$ and a joint $18\times34$ discretization of normalized initial velocity and signed increment.

At $Re_\tau\simeq125$, increment magnitude accounts for approximately $84$--$94\,\%$ of the voxel-level variance of standardized surprisal $\widetilde{S}$, increasing to $92$--$96\,\%$ when the initial state and signed increment are considered jointly. The corresponding joint description accounts for approximately $91$--$97\,\%$ at $Re_\tau\simeq550$. A five-fold temporal hold-out check changes the explained fractions by less than $0.001$ on average relative to the original estimates. This strong relationship is expected: a velocity transition becomes statistically unusual largely because of the magnitude of its observed change, although its conditional likelihood also depends on the initial state.

The population contrasts reflect this relationship. In $5<y^+<12$, the mean-surprisal difference between SHAP-only and background fluid is $1.36$ local standard deviations, and approximately $93\,\%$ of that difference is accounted for by the increment-magnitude reference. This does not eliminate the observed association between surprisal and SHAP-only selection; it establishes that much of the association involves the size of the convected velocity change. The selection analysis below examines whether surprisal retains discriminatory information when increment magnitude is also controlled.

\subsection{Discrimination of SHAP-selected voxels}

The substantial overlap between surprisal and increment magnitude motivates a complementary discrimination test. We compare how well each descriptor distinguishes SHAP-only voxels from unselected fluid outside the classical masks, at similar wall distance and fluctuation energy, using a stratified area under the receiver-operating-characteristic curve (AUC). The AUC is computed by pooling within-stratum pairwise-ranking counts, with ties contributing one half and strata lacking either selected or unselected observations excluded. The tested descriptor is discretized into $50$ percentile bins. An AUC of $0.5$ indicates no discrimination, while values above or below $0.5$ indicate opposite directions of association.

In $5<y^+<12$, surprisal and increment magnitude distinguish SHAP-only selection with AUC values of $0.88$ and $0.85$, respectively [figure~\ref{fig:increment}]. After additionally matching increment magnitude, surprisal retains an AUC of $0.63$. Conversely, after matching surprisal, the AUC of increment magnitude falls to $0.51$. A similar pattern is observed across the examined wall-normal bands. Thus, the two descriptors share much of their association with SHAP selection, but surprisal retains additional discrimination beyond increment magnitude alone.

A separate selection-factor control uses the Mantel--Haenszel estimator in \eqref{eq:mh_selection}, refining each plane--snapshot energy stratum into $50$ equal-frequency bins of $|\Delta U_\ell|$ or $100$ bins of the signed $\Delta U_\ell$. No amplitude normalization is applied in this control, since binning is rank-based within each plane and snapshot. Cells lacking both above- and below-median surprisal do not contribute. In $5<y^+<12$, the signed-increment control retains $48\,\%$ and $41\,\%$ of the SHAP-only volume at $Re_\tau\simeq125$ and $550$, respectively. On these same cells, the energy-only factors are $6.70$ (95\,\% CI $[6.29, 7.11]$) and $2.54$ (95\,\% CI $[2.50, 2.58]$), compared with $7.09$ (95\,\% CI $[6.60, 7.59]$) and $2.67$ (95\,\% CI $[2.62, 2.72]$) after conditioning on the signed increment. The change relative to the unrestricted estimates therefore reflects primarily the reduced support. Confidence intervals use $1000$ bootstrap resamples of blocks spanning $240$ consecutive stored DNS fields ($25$ and $15$ blocks, respectively).

This residual association does not establish an independent informational or dynamical mechanism. At matched energy, SHAP-only selection is also associated with smaller $|u|$ and larger $|w|$, with AUC values of $0.25$ and $0.85$, respectively. These findings characterize the velocity transitions associated with SHAP-only selection without establishing which physical processes determine their predictive importance.

\begin{figure}[!htbp]
    \centering
    \includegraphics[width=0.6\textwidth]{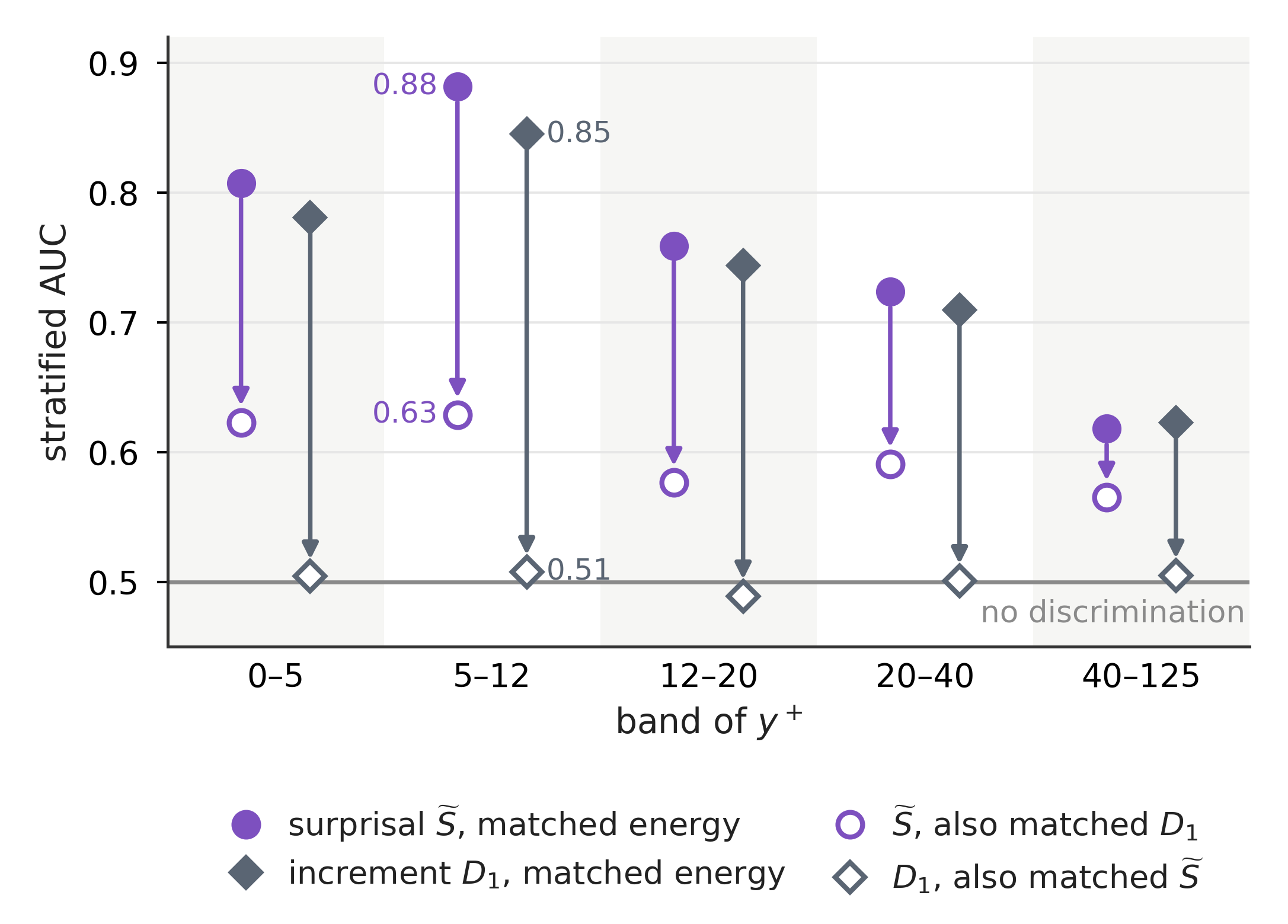}
    \caption{Discrimination of SHAP-only voxels from unselected voxels outside the classical masks using standardized conditional surprisal $\widetilde{S}$ and normalized convected velocity increment $D_1$ at $Re_\tau\simeq125$. Stratified AUC values are shown across wall-normal bands, comparing voxels at similar fluctuation energy and wall distance (filled symbols). Open symbols indicate the corresponding AUC after additionally matching the other descriptor. Circles represent surprisal and diamonds represent increment magnitude; arrows connect the two comparisons for each descriptor. The horizontal line at AUC $=0.5$ indicates no discrimination. Pointwise $95\,\%$ confidence intervals were obtained from $1000$ temporal block-bootstrap replicates using blocks of $120$ consecutive stored DNS fields; their half-widths do not exceed $0.005$ and are mostly obscured by the symbols.}
    \label{fig:increment}
\end{figure}

\subsection{Comparison with generic velocity-sign interfaces}
\label{app:generic_interfaces}

The concentration of surprisal near SHAP-selected HS/LS interfaces is compared with that near other velocity-sign boundaries in the channel, as introduced in \S\ref{sec:pair_interfaces}. Generic reference patches are reconstructed around randomly sampled $u=0$ crossings seeded outside the SHAP mask. For each pair interface, $40$ candidate seeds are tested, constructing patches from crossings within spherical neighbourhoods whose radii are iteratively adjusted to approximate the selected-interface area. The radius is constrained to $2$--$120$ wall units, with at most $12$ iterations and an area tolerance of $25\,\%$. One suitable reference patch is retained per pair when available. Because patches are defined from exterior seeds rather than restricted throughout their extent, some may partially intersect the SHAP mask. A stricter construction excluding such intersections gives the same qualitative result.

The comparison uses $1500$ randomly selected snapshots at $Re_\tau\simeq125$ and $794$ snapshots at $Re_\tau\simeq550$, including both channel halves. Near-surface statistics are evaluated within $|d^+|<2$, using the fraction of voxels with $\widetilde{S}>1.2$. In the direct comparison, this fraction is approximately three times larger near SHAP pair interfaces than near the generic reference surfaces at both Reynolds numbers [figure~\ref{fig:control_u0}(a,b)]. The difference is associated with the substantially greater fluctuation energy of the fluid crossed by the selected interfaces.

To account for differences in fluctuation energy, near-surface voxels are stratified by wall-normal band and by seven classes of relative energy $E/\langle E\rangle(y)$, with class boundaries at $0.25$, $0.5$, $1$, $2$, $4$ and $8$. The strongly surprising voxel fraction at generic surfaces is standardized to the distribution of pair-interface voxels across these strata. Strata without generic reference voxels are excluded, representing less than $0.1\,\%$ of the pair-interface voxel population. The resulting pair-to-generic ratios are $1.16$ (95\,\% CI $[1.09, 1.23]$) and $1.15$ (95\,\% CI $[1.11, 1.20]$) at $Re_\tau\simeq125$ and $550$, respectively [figure~\ref{fig:control_u0}(c)].

A complementary comparison assigns one observation to each reconstructed interface or generic reference patch, rather than weighting individual voxels. Surfaces are stratified by the wall-normal band of the associated pair ($y^+<15$, $15$--$30$, $30$--$60$ and $y^+\geq60$) and the mean relative energy within two wall units of each surface. The generic-surface fractions are standardized to the distribution of pair interfaces across these strata. Only strata with at least five generic reference patches are retained, providing common-support coverage of $77\,\%$ and $91\,\%$ at the two Reynolds numbers. The corresponding ratios are $1.09$ (95\,\% CI $[0.93, 1.27]$) and $1.05$ (95\,\% CI $[0.97, 1.15]$), respectively, both statistically compatible with unity [figure~\ref{fig:control_u0}(c)].

Confidence intervals are percentile intervals obtained from $1000$ temporal block-bootstrap replicates. Blocks of $30$ consecutive snapshots from each chronologically ordered subset are resampled with replacement, preserving the joint sampling of pair interfaces and their generic reference patches. This gives $50$ blocks at $Re_\tau\simeq125$ and $27$ at $Re_\tau\simeq550$, spanning on average approximately $155$ and $123$ stored DNS fields, respectively. The sensitivity of these intervals to longer temporal blocks is discussed in Appendix~\ref{app:bootstrap}.

The residual voxel-weighted excess at matched energy is concentrated mainly near the wall. Interfaces excluded from the surface-level common support tend to be closer to the wall and more energetic than retained interfaces. Nevertheless, restricting the unadjusted comparison to this same support leaves substantial pair-to-generic ratios of $3.06$ and $2.72$ at $Re_\tau\simeq125$ and $550$, respectively. Thus, the pronounced raw contrast is largely associated with the energetic environment of the selected interfaces, rather than being explained by the support restriction alone. A threshold-free comparison gives consistent results: after matching wall distance and energy, the equal-weight-per-surface differences in mean $\widetilde{S}$ are $0.27$ (95\,\% CI $[-0.09, 0.62]$) and $0.02$ (95\,\% CI $[-0.13, 0.19]$), respectively, both statistically compatible with zero. Varying the threshold from $\widetilde{S}>0.8$ to $2.0$ likewise yields energy-matched surface ratios between $1.04$ and $1.09$, with confidence intervals including unity throughout the tested range.

\FloatBarrier
\section{Proximity of SHAP-only voxels to the classical masks}
\label{app:margins}

SHAP-only voxels lie outside the classical coherent-structure masks, but many occur close to their boundaries. We examine whether their association with surprisal can be explained by proximity to classical structures rather than by a distinct statistical property of the selected fluid. Using $1000$ snapshot pairs at $Re_\tau\simeq125$, we compare selected and unselected non-classical voxels at matched wall distance, fluctuation energy, local streamwise-velocity gradient magnitude and distance to the union $\mathcal C$ of classical masks. The gradient descriptor is $|\nabla u|$, with derivatives evaluated spectrally in the homogeneous directions and by finite differences in the wall-normal direction. Matching is performed within wall-normal planes, energy strata, three gradient-magnitude quantile classes and distance intervals bounded by $0$, $2$, $5$, $9$, $13$, $20$, $30$ and $40$ wall units, with a final class for larger distances.

In $5<y^+<12$, approximately $89\,\%$ of SHAP-only volume lies within five wall units of classical fluid, compared with $20\,\%$ of unselected non-classical volume [figure~\ref{fig:margins}(a)]. Much of this difference follows from the energy and gradient distributions: the fraction rises to approximately $77\,\%$ among matched unselected voxels. Moreover, approximately $97\,\%$ of SHAP-only volume belongs to connected SHAP structures containing at least some classically identified fluid. The SHAP-only population should therefore not generally be interpreted as a separate collection of complete coherent structures.

Even after accounting for this proximity, the association between SHAP-only selection and surprisal remains. After matching wall distance, energy, local velocity-gradient magnitude and distance to the classical masks, the near-wall selection factor associated with above-median surprisal remains $8.8$, compared with approximately $10.9$ when only energy and wall distance are controlled [figure~\ref{fig:margins}(b,c)]. The association also persists near the streak threshold: among voxels with streak intensity between $0.8$ and $1$ times this threshold, the selection factor remains $8.0$, using a separate subset of $750$ snapshot pairs. Although the volume classified as SHAP-only depends on the classical thresholds, its association with unusual convected velocity transitions is not explained solely by proximity to classical structures.

\begin{figure}
    \centering
    \includegraphics[width=0.9\textwidth]{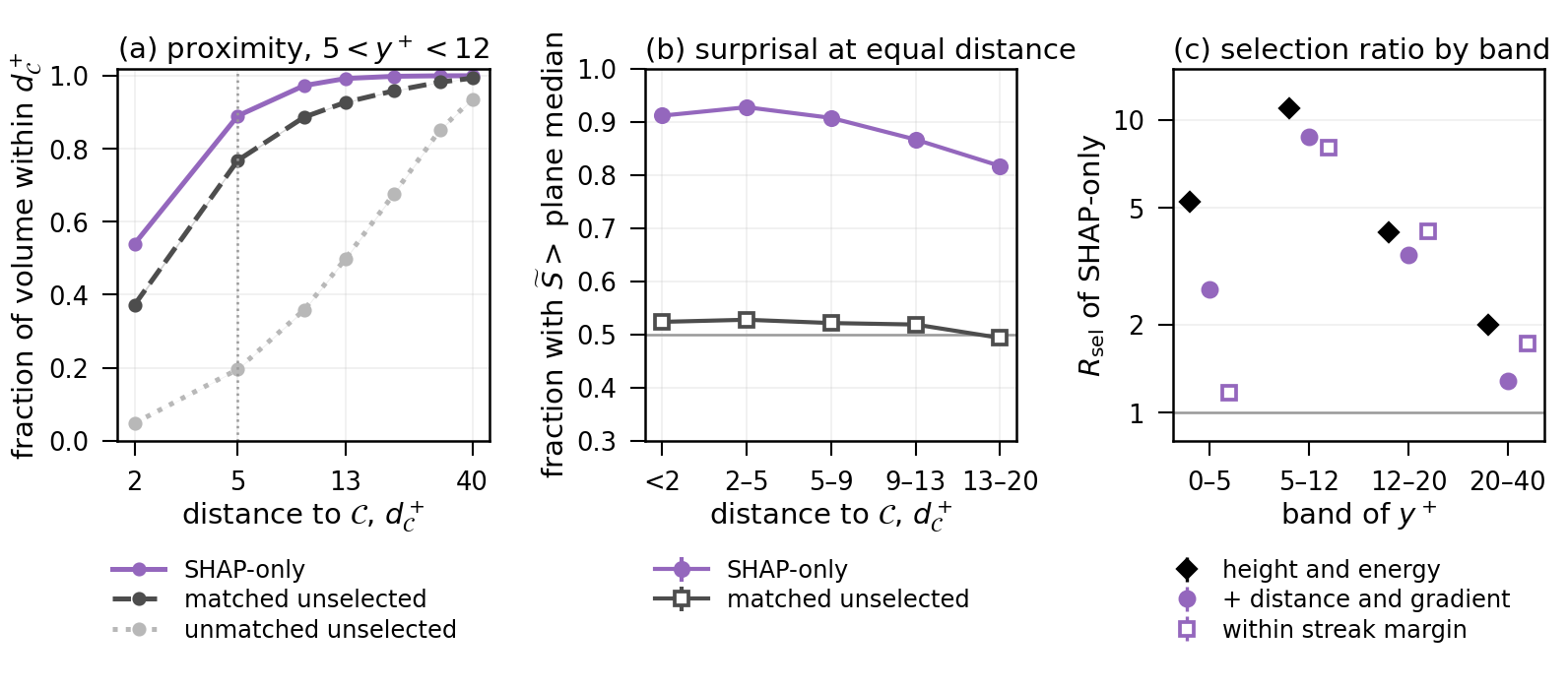}
    \caption{Proximity to classical coherent-structure masks and its association with SHAP-only selection at $Re_\tau\simeq125$. (a) Cumulative voxel fraction against distance to the union $\mathcal C$ for SHAP-only voxels, unselected voxels outside the classical masks and unselected voxels matched in wall distance, fluctuation energy and velocity-gradient magnitude. (b) Fraction of voxels with above-median surprisal in SHAP-only and matched unselected populations, grouped by distance to the classical masks. (c) SHAP-only selection factor under energy-and-height matching, additional distance-and-gradient matching and restriction to near-threshold streak fluid. Error bars indicate $95\,\%$ temporal block-bootstrap confidence intervals.}
    \label{fig:margins}
\end{figure}

\FloatBarrier
\section{Reconstruction and validation of the pair interfaces}
\label{app:interface_composition}

\subsection{Surface reconstruction and classification}

To measure the area and physical composition of the internal HS/LS interfaces in \S\ref{sec:interface_composition}, we reconstruct the $u=0$ surface between the principal high- and low-speed lobes of each SHAP-selected pair as a triangulated mesh. Structures crossing the periodic boundaries are unwrapped before reconstruction. The isosurface is generated using the marching-cubes algorithm of \citet{Lewiner2003}, locating zero crossings by linear interpolation along grid edges and mapping the resulting vertices to physical coordinates to account for the non-uniform wall-normal grid.

Only surface elements associated with contacts between the two principal lobes are retained; interfaces with smaller secondary regions or with the exterior of the SHAP structure are excluded. The surface is partitioned into elements associated with the principal-lobe crossing points, allowing its area and physical composition to be evaluated consistently. Each element is subdivided for numerical integration, and the velocity components, velocity-gradient discriminant and SHAP attribution magnitude are interpolated at the integration points.

The interpolated fields are classified using the same thresholds as in \S\ref{sec:structures}. Surface area is assigned to four mutually exclusive categories: vortical fluid within SHAP structures, $\mathcal H\cap\mathcal V$; non-vortical classical fluid, $\mathcal H\cap(\mathcal Q\cup\mathcal L)\setminus\mathcal V$; SHAP-only portions, $\mathcal H\setminus\mathcal C$; and surface elements classified outside $\mathcal H$. Since $u=0$ on the reconstructed surface, intense Reynolds-stress events contribute negligible area, whereas streaks may intersect the surface through sufficiently large spanwise fluctuations.

The reconstruction uses $1128$ selected snapshots at $Re_\tau\simeq125$ and $794$ at $Re_\tau\simeq550$, retaining only pairs with contacting principal lobes. For uncertainty estimation, the snapshots are assigned to temporal blocks based on consecutive stored DNS field indices, rather than consecutive positions in the analysed subsample. Blocks spanning $120$ stored fields yield $50$ and $29$ non-empty resampling blocks, respectively.

The comparison of interfacial surprisal in figure~\ref{fig:interface_surprisal} uses $16$ area elements per triangular facet, with $\widetilde{S}$ interpolated at each element centre. Elements are assigned to wall-normal bands according to their own heights rather than the pair centroid. For each pair $p$ and band $k$, we compute separate area-weighted mean values $m_{\mathcal V,p,k}$ and $m_{\mathrm{SO},p,k}$ over the vortical and SHAP-only portions. A pair contributes to band $k$ only if each of these two portions occupies at least $5$ squared wall units there. With the associated areas $A_{\mathcal V,p,k}$ and $A_{\mathrm{SO},p,k}$, the between-pair weight is
\begin{equation}
    w_{p,k}=\frac{A_{\mathcal V,p,k}A_{\mathrm{SO},p,k}}{A_{\mathcal V,p,k}+A_{\mathrm{SO},p,k}}.
    \label{eq:interface_pair_weight}
\end{equation}
The category means and their paired difference are then estimated over the same supported pairs:
\begin{equation}
    M_{c,k}=\frac{\sum_p w_{p,k}m_{c,p,k}}{\sum_p w_{p,k}},
    \qquad
    \Delta \widetilde{S}_{\mathcal V-\mathrm{SO},k}
    =\frac{\sum_p w_{p,k}(m_{\mathcal V,p,k}-m_{\mathrm{SO},p,k})}{\sum_p w_{p,k}}
    =M_{\mathcal V,k}-M_{\mathrm{SO},k},
    \label{eq:interface_paired_contrast}
\end{equation}
where $c$ denotes either category. Thus the two means in figure~\ref{fig:interface_surprisal}(a) and their difference in panel (b) use identical pair support and weights; these are not the global area-weighted category means over all interfaces. Confidence intervals are obtained from $2000$ temporal block-bootstrap replicates using blocks of $120$ consecutive stored fields (50 blocks at $Re_\tau\simeq125$ and 29 at $Re_\tau\simeq550$), with the pair-level comparisons resampled jointly within each block. 

To assess how the different energy and increment distributions affect this contrast, a separate control on $200$ snapshots per Reynolds number applies the same paired estimator within bins of local fluctuation energy and convected-increment magnitude. For each surface integration element, $\widetilde{S}$, energy normalized by its plane mean and $D_1$ are interpolated at the element centre. Within each pair and wall-normal band, elements are grouped into $14$ energy bins and $12$ increment-magnitude bins. Only cells containing both vortical and SHAP-only portions contribute, with their area-weighted mean difference weighted by $A_{\mathcal V}A_{\mathrm{SO}}/(A_{\mathcal V}+A_{\mathrm{SO}})$, as in \eqref{eq:interface_pair_weight}. In $12<y^+<20$, the matched cells retain $58\,\%$ and $51\,\%$ of the original paired comparison weight at the two Reynolds numbers. These fractions represent retained comparison weight, not surface area. Confidence intervals are obtained from $1000$ bootstrap resamples of blocks spanning $240$ stored fields, keeping the elements of each pair together. On the retained common support in $12<y^+<20$, the unadjusted paired differences are $-0.08$ (95\,\% CI $[-0.17, 0.00]$) and $-0.14$ (95\,\% CI $[-0.21, -0.06]$) at $Re_\tau\simeq125$ and $550$, respectively. Matching energy and increment magnitude within that support gives $-0.04$ (95\,\% CI $[-0.06, -0.02]$) and $-0.03$ (95\,\% CI $[-0.05, -0.02]$), respectively.

\subsection{Validation and sensitivity}

We assessed sensitivity to the geometric representation and numerical integration of the reconstructed interfaces. At $Re_\tau\simeq125$, an alternative face-based construction gives vortical and SHAP-only area fractions of $38.9\,\%$ and $60.0\,\%$, compared with $41.0\,\%$ and $55.8\,\%$ for the reconstructed surfaces. The difference is largely associated with the treatment of interface edges and the corresponding classification outside the SHAP mask.

In a pilot sample of $11$ pairs, increasing the number of integration elements from $4$ to $64$ changes the area fractions by at most $0.004$, while replacing the marching-cubes implementation of \citet{Lewiner2003} with that of \citet{LorensenCline1987} changes them by at most $0.003$. The sensitivity of the paired vortical-minus-SHAP-only surprisal difference was also examined using alternative quadrature resolutions and marching-cubes implementations in a separate pilot sample at both Reynolds numbers, with changes of at most approximately $0.01$ in standardized-surprisal units. These results support numerical stability under the tested settings, although the limited samples and the treatment of interface edges restrict their generality.

The principal interface accounts for approximately $40\,\%$ of the total contact area between positive- and negative-$u$ regions within the selected pairs. Including contacts associated with smaller secondary lobes changes the vortical and SHAP-only area fractions to $39.9\,\%$ and $55.2\,\%$, respectively. The mixed physical composition reported in \S\ref{sec:interface_composition} is therefore not specific to the restriction to the principal-lobe interface.

\section{Sensitivity to temporal block length}
\label{app:bootstrap}

Confidence intervals throughout the study account for temporal dependence by resampling blocks of consecutive DNS fields. We assess their sensitivity to block length by comparing the reference length of $120$ fields with lengths of $240$ and $480$. The main analyses use approximately $8000$ snapshots at $Re_\tau\simeq125$ and $3400$ at $Re_\tau\simeq550$. The significance of the principal population contrasts, selection factors and interfacial differences remained stable at both Reynolds numbers. Some secondary comparisons, including the LS--HS contrast for streamwise-up pairs at $y^+\geq250$ and $Re_\tau\simeq550$, became sensitive to the block length, as did a few supplementary comparisons in Appendix~\ref{app:generic_interfaces}.

At $Re_\tau\simeq550$, confidence intervals for SHAP-only occupancy widened with longer blocks, by median factors of $1.14$ and $1.38$ for lengths of $240$ and $480$, respectively. Some secondary occupancy statistics also exhibit temporal correlations extending beyond the reference block length. These sensitivities do not change the main near-wall conclusions. The reference block length of $120$ stored DNS fields was therefore retained for the main analysis, with the longer-block comparisons used to assess its limitations.

\section{Sensitivity to structure-identification thresholds}
\label{app:threshold_sensitivity}

The sensitivity to the structure-identification thresholds was assessed by varying each threshold by $\pm10\,\%$, using $200$ snapshots at each Reynolds number. The corresponding masks and voxel populations were reconstructed for each variation. In $5<y^+<12$, the SHAP-only selection factor ranges from $5.9$ to $16.6$ at $Re_\tau\simeq125$ and from $4.1$ to $7.4$ at $Re_\tau\simeq550$, while the corresponding classical-overlap factors remain between approximately $1.1$ and $1.4$. The largest variations arise from the SHAP and streak thresholds, whereas the Q-event and vortical thresholds have a much smaller effect.

We also assess the sensitivity of interfacial composition by varying the SHAP threshold by $\pm10\,\%$ and fully reconstructing the connected structures and interfaces. The resulting SHAP-only interfacial area fractions range from $52.8$ to $59.1\,\%$ at $Re_\tau\simeq125$ and from $52.1$ to $60.9\,\%$ at $Re_\tau\simeq550$. At the higher Reynolds number, lowering the streak threshold by $10\,\%$ reduces this fraction to $50.6\,\%$ (95\,\% CI $[49.2, 52.0]\,\%$). Simultaneously lowering all three classical thresholds by $10\,\%$ gives $49.9\,\%$ (95\,\% CI $[48.5, 51.3]\,\%$); SHAP-only remains the largest interfacial category but is not a statistically established majority under this perturbation.

\end{appen}

\clearpage

\end{document}